\documentclass[twocolumn,aps,prl,showpacs,groupedaddress,amsfonts,amssymb,amsmath]{revtex4-1}
\usepackage{graphicx}
\usepackage{xcolor, soul}
\usepackage{subfigure}
\usepackage{epstopdf}
\usepackage[colorlinks=true, letterpaper=true, pdfstartview=FitV, linkcolor=blue, citecolor=blue, urlcolor=blue]{hyperref}
\begin{document}
\draft

\title{Topological Anderson Amorphous Insulator}

\author{Xiaoyu Cheng$^{1,3}$, Tiantao Qu$^{2}$, Liantuan Xiao$^{1,3}$, Suotang Jia$^{1,3}$, Jun Chen$^{2,3,\dagger}$, and Lei Zhang$^{1,3,\ddagger}$}
\address{$^1$State Key Laboratory of Quantum Optics and Quantum Optics Devices, Institute of Laser Spectroscopy, Shanxi University, Taiyuan 030006, China\\
$^2$State Key Laboratory of Quantum Optics and Quantum Optics Devices, Institute of Theoretical Physics, Shanxi University, Taiyuan 030006, China\\
$^3$Collaborative Innovation Center of Extreme Optics, Shanxi University, Taiyuan 030006, China}

\begin{abstract}
The topological phase in amorphous systems adds a new dimension to the topological states of matter. Here, we present an interesting phenomenon dubbed the topological Anderson amorphous insulator (TAAI). Anderson disorder can drive topologically trivial amorphous systems with structural disorders into non-crystalline topological insulators. The gap closing and reopening, spin Bott index, robust edge states, and quantized conductance characterize the Anderson disorder-induced nontrivial topology in amorphous systems. More importantly, phase diagrams are given for the topological phase transition (TPT). It is found that amorphous structural disorder and Anderson disorder are synergistic to drive the $s-p$ band inversion of the system and hence the TPT, which is further confirmed by the effective medium theory. Our findings report a disorder-induced topological phenomenon in non-crystalline systems and shed light on the physical understanding of the interplay between the coexistence of two types of disorder effects and topology.
\end{abstract}

\maketitle

\emph{Introduction}-As one of the intriguing phenomena in the interplay between topology and disorders of topological systems, the topological Anderson insulator (TAI) attracts much research attention\cite{li2009topological,groth2009theory,jiang2009numerical,prodan2011three,guo2010topological,loring2011disordered,xing2011topological,fulga2011scattering,akhmerov2011quantized,ryu2012disorder,song2012dependence,altland2014quantum,liu2017disorder,meier2018observation,stutzer2018photonic,li2020topological,zhang2020non,liu2020topological,cui2022photonic,liu2021real,yang2021higher,shi2021disorder,lin2022observation,chen2019topological}. Typically, the topological system becomes a trivial insulator, and all states become localized when the disorder strength is sufficiently large due to the Anderson localization. However, TAI refers to the Anderson disorder-induced topological phase transition initially from the trivial phase. It is proposed in various systems, such as the HgTe/CdTe quantum well\cite{li2009topological,groth2009theory,jiang2009numerical,prodan2011three}, the Su-Schrieffer-Heeger (SSH) chain\cite{su1979solitons}, etc. Until now, TAI phenomena have been experimentally confirmed in one-dimensional cold atomic wires\cite{meier2018observation} and two-dimensional (2D) photonic waveguide array systems\cite{stutzer2018photonic}.

As well known, another disorder mechanism different from Anderson disorder is the amorphous structural disorder. In nature, many materials exist in the amorphous phase, for instance, glasses, polymers, and gels. Amorphous structural disorder means that there is no regularly arranged lattice structure in the material but a disordered arrangement similar to glass, which usually reduces and destroys the periodicity and symmetry of the system and even eliminates the energy band topology of electrons\cite{weaire1971electronic,zallen2008physics,yang2021determining,focassio2021structural}. Recently, the concept of amorphous topological insulators has been theoretically proposed in condensed matter systems, photonic crystals, etc.\cite{rechtsman2011amorphous,mitchell2018amorphous,poyhonen2018amorphous,agarwala2019topological,mansha2017robust,costa2019toward,agarwala2020higher,spring2021amorphous,focassio2021amorphous,yang2019topological,ivaki2020criticality,wang2022structural,peng2022density,wang2021structural,zhang2023anomalous}, and experimentally realized\cite{zhou2020photonic,corbae2023observation}. More interestingly, the amorphous structural disorder induces topological transition is predicted\cite{wang2022structural}.

Physically, Anderson disorder and structural disorder are two different disorder mechanisms that have different effects on the electronic properties of materials. The amorphous structural disorder is rooted in the renormalization of the spectral gap in association with single-particle energy levels, while Anderson disorder is originally a many-body effect. Naturally, the following questions arise, can Anderson disorder-induced topological states in a topologically trivial amorphous system? When Anderson disorder and amorphous structural disorder coexist, are the two mechanisms competing or synergizing with each other with the interplay of topology?

In this work, we report an Anderson disorder-induced topological phase transition (TPT) starting from a trivial amorphous system with tight-binding simulations, dubbed topological Anderson amorphous insulator (TAAI). This exciting phenomenon is confirmed by investigating the gap closing and reopening process accompanied by an abrupt change in topological invariant spin Bott index $B_s$. Furthermore, we calculate the real space distribution of eigenstates in amorphous systems and quantized conductance to verify the existence of robust edge states in amorphous systems with Anderson disorder. Topological phase diagrams indicate a large parameter space for TAAI realization, showing the synergistic mechanism of two types of disorders. Our study provides a more comprehensive understanding of how two disorder mechanisms cooperate to give rise to the TAAI phase.
\begin{figure*}
\centering
\includegraphics[scale=0.54]{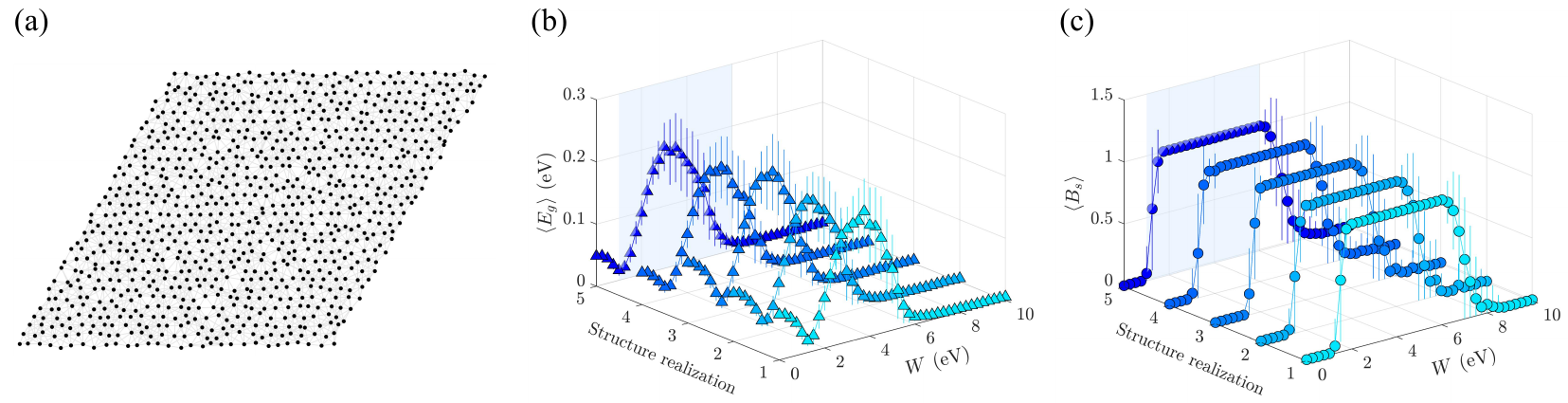}
\caption{(a) Schematic diagram of an amorphous system with 900 atoms and their corresponding bondings ($\sigma=0.15$). (b,c) Disorder averaged bulk energy gap $\langle E_g \rangle$ and spin Bott index $\langle B_s \rangle$ as functions of Anderson disorder strength $W$ for five realizations of amorphous system, respectively. The parameters are $\epsilon_s = 1.8$ eV, $\epsilon_p= -6.5$ eV, $V_{ss\sigma} = -0.256$ eV, $V_{sp\sigma} = 0.576$ eV, $V_{pp\sigma} = 1.152$ eV, $V_{pp\pi}= 0.032$ eV, and $\lambda = 0.8$ eV. Each data point on the figure is averaged over 100 disorder configurations.}
\label{Fig1}
\end{figure*}

\emph{Model Hamiltonian}-Since the amorphization process is rather complicated, many theoretical models are constructed to simulate amorphous lattice\cite{zallen2008physics,wooten1985computer,car1988structural,tu1998properties,hannemann2004modeling,drabold2009topics,youn2014efficient}. Here, the thermal fluctuations of atoms corresponding to a typical melting-quenching process is adopted to construct the desired amorphous system\cite{duwez1960continuous,wang2022structural}. It is assumed that a random displacement $\mathbf{r}$ is assigned to each site of an original 2D triangular lattice and the atomic displacements follow a Normal distribution $N(\mu,\sigma^2)$ in random directions shown in Fig. \ref{Fig1}(a). Physically, atoms are approximately kept in their equilibrium positions by harmonic forces\cite{koski1961non}. For simplicity, we take $\mu=0$, and the standard deviation $\sigma$ denotes the strength of the system's amorphous structural disorder.

We start from an atomic-based tight-binding model Hamiltonian for quasi-lattice of amorphous system\cite{wang2016quantum,huang2018quantum,wang2022structural}
\begin{equation}
\begin{aligned}
H
&= \sum_{i\alpha}\epsilon_{\alpha}c_{i\alpha}^{\dagger}c_{i\alpha}+\sum_{\langle i\alpha,j\beta\rangle}t_{\alpha,\beta}(\mathbf{r}_{ij})c_{i\alpha}^{\dagger}c_{j\beta}\\
&+i\lambda\sum_{i}(c_{ip_{y}}^{\dagger}\sigma_{z}c_{ip_{x}}-c_{ip_{x}}^{\dagger}\sigma_{z}c_{ip_{y}}),\label{HamH}
\end{aligned}
\end{equation}
where $c_{i\alpha}^{\dagger}=(c_{i\alpha\uparrow}^{\dagger},c_{i\alpha\downarrow}^{\dagger})$ represents the creation operators at site $i$ with $\alpha$ denoting three orbitals $(s,p_x,p_y)$ in each site. The first term describes the on-site energy of $\alpha$ orbital. In addition, static Anderson disorder is added to $\epsilon_{\alpha}$ with a uniform distribution in the interval $[-W/2, W/2]$ where $W$ characterizes the strength of the Anderson disorder. The second term is the hopping integral term $t_{\alpha,\beta}(\mathbf{r}_{ij})$, which is given by
\begin{equation}
t_{\alpha,\beta}(\mathbf{r}_{ij}) =
\frac{\Theta(R-r_{ij})}{r_{ij}^{2}}\rm{SK}
\begin{bmatrix}
V_{\alpha\beta\delta},\hat{\mathbf{r}}_{ij}
\end{bmatrix}.\label{HamT}
\end{equation}
It depends on the orbital type ($\alpha$ and $\beta$), the intersite vector $\mathbf{r}_{ij}$, and $\rm{SK}[\cdot]$ represents Slater-Koster parametrization for three  orbitals\cite{slater1954simplified}, $V_{\alpha\beta\delta}$ $(\delta= \sigma $ or $ \pi)$ are the bond parameters, and $\hat{\mathbf{r}}_{ij}$ is the unit direction vector, $\Theta(R-r_{ij})$ is the step function with cut-off parameter $R=1.9a$. The distance dependence of the bonding parameters is assumed to be approximately captured by the Harrison relation\cite{harrison2012electronic} and $r_{ij}$ is set to a constant $2r_{0}$ with $r_0=0.2$. Note that hoppings for $\mid\mathbf{r}_{ij}\mid>R$ are neglected in the numerical simulation, and the unit of lattice constant $a=1$ is chosen for simplicity. Here, the topological transition relies on the band inversion between $s$ and $p$ states, and the $2/3$ filling of electron states is chosen in the following. The third term in Eq. (\ref{HamH}) is the spin-orbital interaction, and $\lambda$ denotes its strength.

\emph{Finite systems}-Figure. \ref{Fig1}(a) illustrates an example of 2D amorphous system with $900$ atoms, which is randomly selected from the distribution with $\sigma=0.15$. Once the amorphous structure is fixed, i.e., a specific realization, we can study the Anderson disorder effect on it. Here, the periodic boundary condition (PBC) is used and five realizations are randomly selected. The disorder averaged bulk energy gap $\langle E_g \rangle$ as a function of the Anderson disorder strength $W$ is presented in Fig. \ref{Fig1}(b). Error bars show fluctuations under different disorder configurations in the same realization. It is found that, as the disorder strength increases up to around $1.5$ eV, the energy gap first drops to nearly zero and then starts to increase. This energy gap closing and reopening indicates the occurrence of TPT. To verify whether it is TPT or not, we further calculate the topological invariant in real space, the spin Bott index $B_s$, which can characterize quantum spin Hall (QSH) effect phase in finite systems\cite{huang2018quantum,huang2018theory}. As shown in Fig. \ref{Fig1}(c), the disorder averaged spin Bott index $\langle B_s \rangle$ is 0 before the energy gap closing. As the disorder strength $W$ increases, the $\langle B_s \rangle$ jumps sharply from 0 to 1 when passing through the energy gap closing phase transition point in Fig. \ref{Fig1}(b). The $\langle B_s \rangle$ remains at 1 over a wide range of disorder strength $W \in [2,6]$ eV without any fluctuations. This confirms that Anderson disorder indeed induces the TPT from a trivial amorphous system, which is named as TAAI. Note that the same physics happens in all realizations, indicating that the discovered TAAI phenomenon is robust and independent of realizations. The system's energy gap eventually closes when the disorder strength is strong enough, which enters the Anderson localization regime.

\begin{figure}
\centering
\includegraphics[scale=0.35]{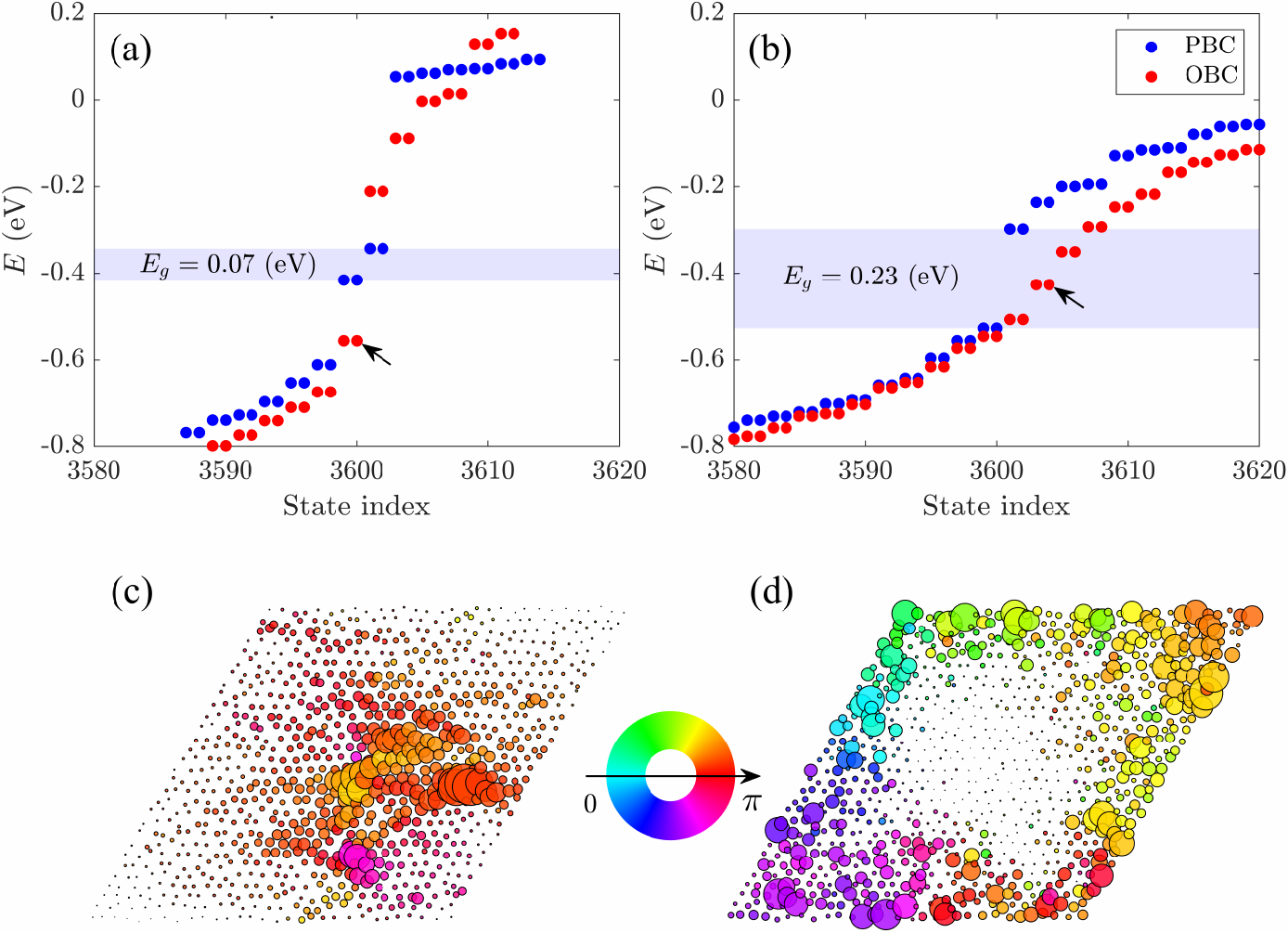}
\caption{(a) Energy eigenvalues $E$ versus the state index for an amorphous system with 900 atoms ($\sigma=0.15, W=0$ eV) under PBC and OBC. (b) Energy eigenvalues of the same amorphous realization in (a) within the presence of a specific Anderson disorder configuration ($W=4$ eV). (c,d) The wave function distribution ($p$ and $s$ orbitals) in real space corresponding to the eigenstates indicated by the black arrows in (a) $E=-0.55$ eV of the clean system and (b) $E=-0.43$ eV with disorders, respectively. The size and color of the spots represent the amplitude and phase of the wave function, respectively. The parameters are identical to those in Fig. \ref{Fig1}.}
\label{Fig2}
\end{figure}

To show the topological edge states induced by Anderson disorder in the amorphous system more clearly, we calculate the energy eigenvalue versus the state index of a clean amorphous system ($W=0$ eV) with PBC and open boundary condition (OBC). As shown in Fig. \ref{Fig2}(a), the system has an energy gap of 0.07 eV. The energy gap increases when the system is under OBC, indicating that the amorphous system is an insulator. Figure. \ref{Fig2}(c) shows the wave function distribution of $p$ orbital corresponding to the eigenstate when the energy $E=-0.55$ eV under the OBC. It is seen that the wave function is mainly distributed the central region of the system. When the Anderson disorder with strength $W=4$ eV is introduced to the same structural realization, the energy eigenvalue versus the state index under the PBC and OBC are presented in Fig. \ref{Fig2}(b). Due to the presence of Anderson disorder, the energy gap is 0.23 eV now under the PBC. Interestingly, when the system is in OBC, a series of eigenstates show up in the energy gap compared to the gap increasing in the clean case. This also partially reflects that the system undergoes a transition from a trivial insulating phase to a topological insulator phase. Likewise, the wave function distribution of $s$ orbital corresponding to energy $E=-0.43$ eV is shown in Fig. \ref{Fig2}(d). In contrast to Fig. \ref{Fig2}(c), these eigenstates in the energy gap are mainly distributed on the system's boundary, i.e., edge states. This further verifies the topological properties of TAAI system with $\langle B_s \rangle=1$. Note that the spin $S_z$ symmetry is preserved since we consider a nonmagnetic Anderson disorder with time-reversal symmetry.  We also present the $s-p$ band inversion process of Anderson disorder-induced TAAI via wave functions of three orbitals (see Supplemental Material\cite{suppsee}).

\begin{figure}
\centering
\includegraphics[scale=0.50]{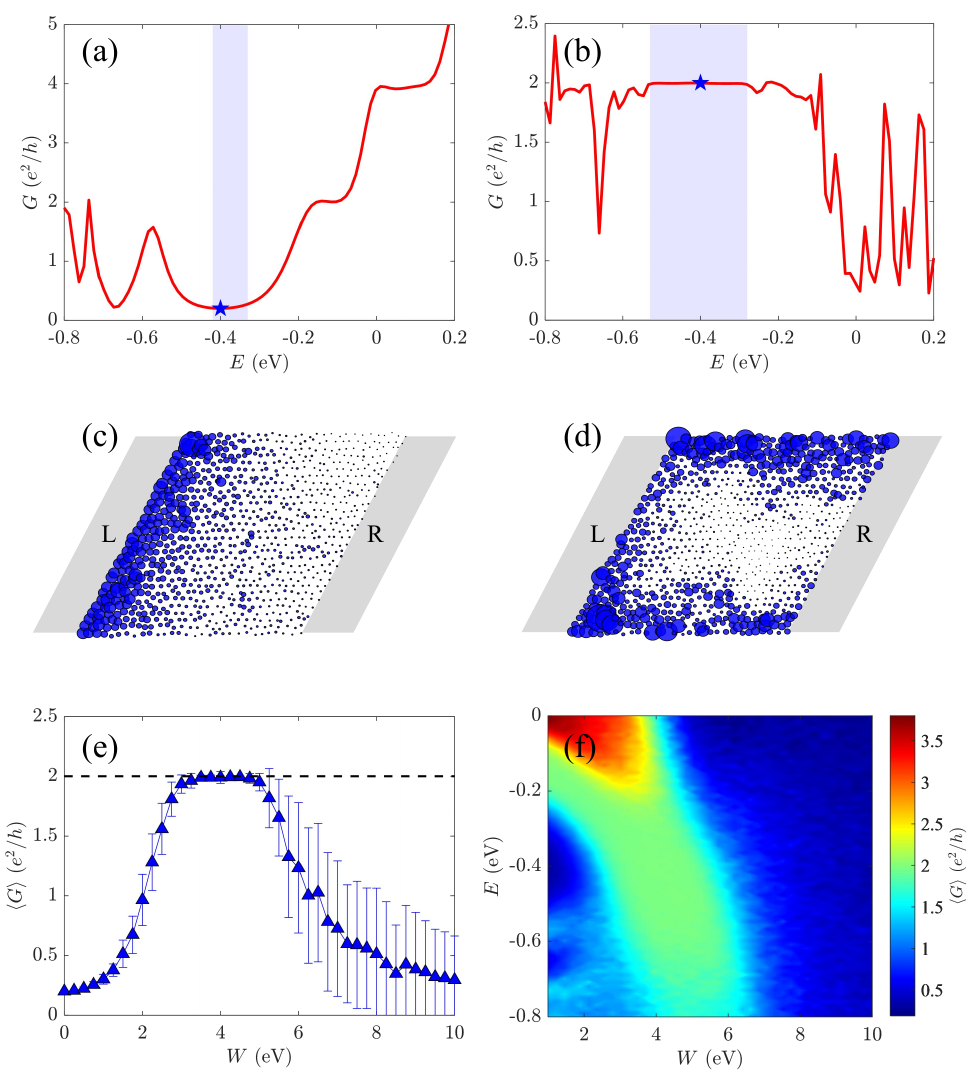}
\caption{Quantum transport calculations in an amorphous system contacted with two leads. (a) Conductance $G$ versus electron energy in a clean system ($W=0$ eV). (b) Conductance $G$ versus electron energy with Anderson disorder strength $W=4$ eV. (c,d) The local density of scattering states (LDOSS) injected from left lead in real space when the electron energy $E=-0.4$ eV marked by the blue star in (a) and (b), respectively. Light gray denotes the left and right metallic leads. (e) The disorder averaged conductance $\langle G \rangle$ versus Anderson disorder strength $W$. (f) Phase diagram of $\langle G \rangle$ versus the electron energy $E$ and Anderson disorder strength $W$. Colors represent the magnitude of averaged conductance. Each data point on the figure is averaged over $100$ disorder configurations.}
\label{Fig3}
\end{figure}

\emph{Open systems}-Since the topological edge states can lead to the quantized conductance as a signature, we further study the transport properties of edge states of TAAI. By connecting two semi-infinite leads, the conductance of the amorphous system with or without Anderson disorder can be obtained based on the non-equilibrium Green's function method\cite{datta1997electronic,khomyakov2005conductance,wang2009relation,xing2011topological,zhang2012first,zhang2014universal,cheng2021antichiral}. Figure. \ref{Fig3}(a) shows the conductance $G$ in a clean amorphous system versus the electron energy. It can be seen that when the electron energy locates in the energy gap (the shaded region corresponds to the energy gap in Fig. \ref{Fig2}(a)), the conductance $G$ is quite small, indicating that the system is a normal insulator. Correspondingly, Fig. \ref{Fig3}(c) presents the local density of scattering states (LDOSS) injected from the left lead of energy marked by a blue star in Fig. \ref{Fig3}(a). The LDOSS distributes around the left lead and no transport channel is nearly formed in the amorphous system. In contrast, when Anderson disorder is added, compared to Fig. \ref{Fig3}(a), a quantized platform with conductance $G=2e^{2}/h$ appears in the energy gap region (the shaded region corresponds to the energy gap in Fig. \ref{Fig2}(b)). At the same time, the LDOSS is mainly distributed on the upper and lower boundaries of the system shown in Fig. \ref{Fig3}(d). In fact, two transport channels carry spin-up and spin-down components respectively, which give rise to the quantized conductance. When the electron energy is fixed as $E=-0.4$ eV, we calculate the Anderson disorder averaged conductance $\langle G \rangle$ versus the $W$. From Fig. \ref{Fig3}(e), we can see that the quantized conductance plateau can be induced in a wide range of disorder strengths $W \in [3,5]$ eV, accompanying zero fluctuation. Figure. \ref{Fig3}(f) presents the phase diagram of the $\langle G \rangle$ in the parameter space of electron energy and Anderson disorder strength. The quantized conductance denoted with green color occupies a large area of the parameter space. These quantized signatures provide an excellent operating room for detecting the TAAI phase.

\begin{figure}
\centering
\includegraphics[scale=0.65]{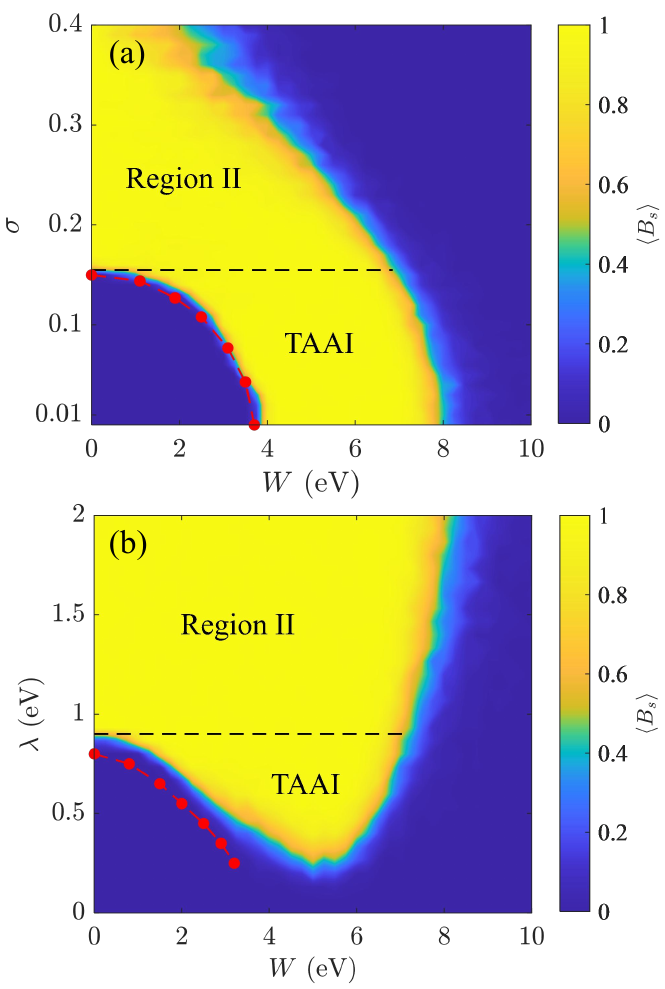}
\caption{(a) Phase diagram of the averaged spin Bott index $\langle B_s \rangle$ in the parameter space of the amorphous structural disorder strength $\sigma$ and the Anderson disorder strength $W$ with $\lambda=0.8$ eV. (b) Phase diagram of the $\langle B_s \rangle$ in the parameter space of the SOC strength $\lambda$ and $W$ with $\sigma=0.15$. Red dashed dot lines in (a,b) are the averaged energy gap closing points $\langle E_g\rangle=0$ obtained by the effective medium theory. Every data point on the figure is averaged over ten realizations of amorphous structures with $20$ Anderson disorder configurations in each.}
\label{Fig4}
\end{figure}

\emph{Phase diagrams}-Above, TAAI is proved to exist with a specific structural disorder strength $\sigma=0.15$. In the following, we will show that TAAI resides in a wide range of $\sigma$ instead of an accidental case. The averaged spin Bott index $\langle B_s \rangle$ is calculated by averaging over ten realizations of amorphous structures with 20 Anderson disorder configurations in each. By computing the phase diagram of the $\langle B_s \rangle$ in the parameter space of the $\sigma$ and the $W$, we can find a continuous yellow area corresponds to a quantized $\langle B_s \rangle=1$ as shown in Fig. \ref{Fig4}(a). Note that the yellow region can be separated into two regions by the horizontal dashed line in the figure. The TAAI phase can be formed when $\sigma$ is smaller than $\sigma_c=0.155$, which confirms its robustness. It is easy to find that when $\sigma$ is small, a large $W$ is required to induce the topological states. When $\sigma$ increases, only a small $W$ is required. These two disorder mechanisms synergize to drive the system to the TAAI phase from the trivial phase. Furthermore, it can be found that the $\langle B_s \rangle$ is already 1 in the clean limit when the $\sigma$ is larger than the critical value $\sigma_c$. The structural disorder already induces the non-trivial topological insulator phase in Region II even when the Anderson disorder is absent. As the Anderson disorder strength increases, the system first persists in the topological insulator phase and finally becomes a normal insulator due to the Anderson localization. This indicates that the amorphous system is in a topological insulator phase denoting Region II in Fig. \ref{Fig4}(a).

Since the SOC strength $\lambda$ is critically important of realizing the QSH states, the phase diagram of the $\langle B_s \rangle$ in the parameter space of the $\lambda$ and the $W$ is also calculated with $\sigma=0.15$ and presented in Fig. \ref{Fig4}(b). It is found that there exists a wide range for the parameter $\lambda\in[0.25, 0.9]$ eV. When $W$ is rather small, the system is in the trivial phase. As $W$ increases, the system can pass through the phase boundary and suddenly change from the trivial phase to the TAAI. When $\lambda$ is large enough, the system is already in the topologically nontrivial phase even when there is no Anderson disorder involved. It is worth mentioning that the TAAI can be realized around the phase boundary with smaller Anderson disorder strength required, for instance, $\lambda=0.88$ eV, $W=0.95$ eV, which is a more accessible parameter in the experiment.

To further confirm the emergence of TAAI, we calculate the averaged energy gap closing points $\langle E_g\rangle=0$ with the effective medium theory (EMT). The amorphous structural disorder and Anderson disorder are treated as effective self-energy to renormalize the system's energies, which induce the band inversion. The results are shown as the red dashed dot line in Figs. \ref{Fig4}(a,b). It can be seen that the phase boundary of TPT in $\sigma-W$ and $\lambda-W$ planes can be well captured by EMT, which further confirms the existence of TAAI. The EMT can explain the synergistic effect of structural and Anderson disorders during the TPT. Detailed information on EMT is presented in the Supplemental Material\cite{suppsee}.

\emph{Conclusion}-We report an interesting Anderson disorder-induced TPT phenomenon, i.e., a trivial amorphous insulator to a TAAI. This TPT is confirmed by the gap-closing and gap-reopening process accompanied by a sudden jump of the averaged spin Bott index. Furthermore, to verify its existence, Anderson disorder-induced topological edge states and the disorder averaged quantized conductance $\langle G \rangle$ are presented. The topological phase diagrams confirm the robustness of TAAI and provide valuable information in parameters' selections to realize TAAI. Moreover, it is found that the effective medium theory can capture the TPT. Since the topological amorphous systems have been experimentally demonstrated in photonic crystal\cite{zhou2020photonic} and topological circuit systems\cite{yang2019topological}, we expect the TAAI phenomenon to be also realized on these platforms.

We gratefully acknowledge the support from the National Key R\&D Program of China under Grant No. 2022YFA1404003, the National Natural Science Foundation of China (Grant No. 12074230, 12174231), the Fund for Shanxi ``1331 Project", Shanxi Province 100-Plan Talent Program, Fundamental Research Program of Shanxi Province through 202103021222001, Research Project Supported by Shanxi Scholarship Council of China. This research was partially conducted using the High Performance Computer of Shanxi University.

\bigskip
\noindent{$^{\dagger}$chenjun@sxu.edu.cn}\\
\noindent{$^{\ddagger)}$zhanglei@sxu.edu.cn}

\bibliography{amorphous}

\begin{thebibliography}{69}%
\makeatletter
\providecommand \@ifxundefined [1]{%
 \@ifx{#1\undefined}
}%
\providecommand \@ifnum [1]{%
 \ifnum #1\expandafter \@firstoftwo
 \else \expandafter \@secondoftwo
 \fi
}%
\providecommand \@ifx [1]{%
 \ifx #1\expandafter \@firstoftwo
 \else \expandafter \@secondoftwo
 \fi
}%
\providecommand \natexlab [1]{#1}%
\providecommand \enquote  [1]{``#1''}%
\providecommand \bibnamefont  [1]{#1}%
\providecommand \bibfnamefont [1]{#1}%
\providecommand \citenamefont [1]{#1}%
\providecommand \href@noop [0]{\@secondoftwo}%
\providecommand \href [0]{\begingroup \@sanitize@url \@href}%
\providecommand \@href[1]{\@@startlink{#1}\@@href}%
\providecommand \@@href[1]{\endgroup#1\@@endlink}%
\providecommand \@sanitize@url [0]{\catcode `\\12\catcode `\$12\catcode
  `\&12\catcode `\#12\catcode `\^12\catcode `\_12\catcode `\%12\relax}%
\providecommand \@@startlink[1]{}%
\providecommand \@@endlink[0]{}%
\providecommand \url  [0]{\begingroup\@sanitize@url \@url }%
\providecommand \@url [1]{\endgroup\@href {#1}{\urlprefix }}%
\providecommand \urlprefix  [0]{URL }%
\providecommand \Eprint [0]{\href }%
\providecommand \doibase [0]{http://dx.doi.org/}%
\providecommand \selectlanguage [0]{\@gobble}%
\providecommand \bibinfo  [0]{\@secondoftwo}%
\providecommand \bibfield  [0]{\@secondoftwo}%
\providecommand \translation [1]{[#1]}%
\providecommand \BibitemOpen [0]{}%
\providecommand \bibitemStop [0]{}%
\providecommand \bibitemNoStop [0]{.\EOS\space}%
\providecommand \EOS [0]{\spacefactor3000\relax}%
\providecommand \BibitemShut  [1]{\csname bibitem#1\endcsname}%
\let\auto@bib@innerbib\@empty
\bibitem [{\citenamefont {Li}\ \emph {et~al.}(2009)\citenamefont {Li},
  \citenamefont {Chu}, \citenamefont {Jain},\ and\ \citenamefont
  {Shen}}]{li2009topological}%
  \BibitemOpen
  \bibfield  {author} {\bibinfo {author} {\bibfnamefont {J.}~\bibnamefont
  {Li}}, \bibinfo {author} {\bibfnamefont {R.-L.}\ \bibnamefont {Chu}},
  \bibinfo {author} {\bibfnamefont {J.~K.}\ \bibnamefont {Jain}}, \ and\
  \bibinfo {author} {\bibfnamefont {S.-Q.}\ \bibnamefont {Shen}},\ }\href@noop
  {} {\bibfield  {journal} {\bibinfo  {journal} {Phys. Rev. Lett.}\ }\textbf
  {\bibinfo {volume} {102}},\ \bibinfo {pages} {136806} (\bibinfo {year}
  {2009})}\BibitemShut {NoStop}%
\bibitem [{\citenamefont {Groth}\ \emph {et~al.}(2009)\citenamefont {Groth},
  \citenamefont {Wimmer}, \citenamefont {Akhmerov}, \citenamefont
  {Tworzyd{\l}o},\ and\ \citenamefont {Beenakker}}]{groth2009theory}%
  \BibitemOpen
  \bibfield  {author} {\bibinfo {author} {\bibfnamefont {C.~W.}\ \bibnamefont
  {Groth}}, \bibinfo {author} {\bibfnamefont {M.}~\bibnamefont {Wimmer}},
  \bibinfo {author} {\bibfnamefont {A.~R.}\ \bibnamefont {Akhmerov}}, \bibinfo
  {author} {\bibfnamefont {J.}~\bibnamefont {Tworzyd{\l}o}}, \ and\ \bibinfo
  {author} {\bibfnamefont {C.~W.~J.}\ \bibnamefont {Beenakker}},\ }\href@noop
  {} {\bibfield  {journal} {\bibinfo  {journal} {Phys. Rev. Lett.}\ }\textbf
  {\bibinfo {volume} {103}},\ \bibinfo {pages} {196805} (\bibinfo {year}
  {2009})}\BibitemShut {NoStop}%
\bibitem [{\citenamefont {Jiang}\ \emph {et~al.}(2009)\citenamefont {Jiang},
  \citenamefont {Wang}, \citenamefont {Sun},\ and\ \citenamefont
  {Xie}}]{jiang2009numerical}%
  \BibitemOpen
  \bibfield  {author} {\bibinfo {author} {\bibfnamefont {H.}~\bibnamefont
  {Jiang}}, \bibinfo {author} {\bibfnamefont {L.}~\bibnamefont {Wang}},
  \bibinfo {author} {\bibfnamefont {Q.-f.}\ \bibnamefont {Sun}}, \ and\
  \bibinfo {author} {\bibfnamefont {X.~C.}\ \bibnamefont {Xie}},\ }\href@noop
  {} {\bibfield  {journal} {\bibinfo  {journal} {Phys. Rev. B}\ }\textbf
  {\bibinfo {volume} {80}},\ \bibinfo {pages} {165316} (\bibinfo {year}
  {2009})}\BibitemShut {NoStop}%
\bibitem [{\citenamefont {Prodan}(2011)}]{prodan2011three}%
  \BibitemOpen
  \bibfield  {author} {\bibinfo {author} {\bibfnamefont {E.}~\bibnamefont
  {Prodan}},\ }\href@noop {} {\bibfield  {journal} {\bibinfo  {journal} {Phys.
  Rev. B}\ }\textbf {\bibinfo {volume} {83}},\ \bibinfo {pages} {195119}
  (\bibinfo {year} {2011})}\BibitemShut {NoStop}%
\bibitem [{\citenamefont {Guo}\ \emph {et~al.}(2010)\citenamefont {Guo},
  \citenamefont {Rosenberg}, \citenamefont {Refael},\ and\ \citenamefont
  {Franz}}]{guo2010topological}%
  \BibitemOpen
  \bibfield  {author} {\bibinfo {author} {\bibfnamefont {H.-M.}\ \bibnamefont
  {Guo}}, \bibinfo {author} {\bibfnamefont {G.}~\bibnamefont {Rosenberg}},
  \bibinfo {author} {\bibfnamefont {G.}~\bibnamefont {Refael}}, \ and\ \bibinfo
  {author} {\bibfnamefont {M.}~\bibnamefont {Franz}},\ }\href@noop {}
  {\bibfield  {journal} {\bibinfo  {journal} {Phys. Rev. Lett.}\ }\textbf
  {\bibinfo {volume} {105}},\ \bibinfo {pages} {216601} (\bibinfo {year}
  {2010})}\BibitemShut {NoStop}%
\bibitem [{\citenamefont {Loring}\ and\ \citenamefont
  {Hastings}(2010)}]{loring2011disordered}%
  \BibitemOpen
  \bibfield  {author} {\bibinfo {author} {\bibfnamefont {T.~A.}\ \bibnamefont
  {Loring}}\ and\ \bibinfo {author} {\bibfnamefont {M.~B.}\ \bibnamefont
  {Hastings}},\ }\href@noop {} {\bibfield  {journal} {\bibinfo  {journal}
  {Europhys. Lett.}\ }\textbf {\bibinfo {volume} {92}},\ \bibinfo {pages}
  {67004} (\bibinfo {year} {2010})}\BibitemShut {NoStop}%
\bibitem [{\citenamefont {Xing}\ \emph {et~al.}(2011)\citenamefont {Xing},
  \citenamefont {Zhang},\ and\ \citenamefont {Wang}}]{xing2011topological}%
  \BibitemOpen
  \bibfield  {author} {\bibinfo {author} {\bibfnamefont {Y.}~\bibnamefont
  {Xing}}, \bibinfo {author} {\bibfnamefont {L.}~\bibnamefont {Zhang}}, \ and\
  \bibinfo {author} {\bibfnamefont {J.}~\bibnamefont {Wang}},\ }\href@noop {}
  {\bibfield  {journal} {\bibinfo  {journal} {Phys. Rev. B}\ }\textbf {\bibinfo
  {volume} {84}},\ \bibinfo {pages} {035110} (\bibinfo {year}
  {2011})}\BibitemShut {NoStop}%
\bibitem [{\citenamefont {Fulga}\ \emph {et~al.}(2011)\citenamefont {Fulga},
  \citenamefont {Hassler}, \citenamefont {Akhmerov},\ and\ \citenamefont
  {Beenakker}}]{fulga2011scattering}%
  \BibitemOpen
  \bibfield  {author} {\bibinfo {author} {\bibfnamefont {I.~C.}\ \bibnamefont
  {Fulga}}, \bibinfo {author} {\bibfnamefont {F.}~\bibnamefont {Hassler}},
  \bibinfo {author} {\bibfnamefont {A.~R.}\ \bibnamefont {Akhmerov}}, \ and\
  \bibinfo {author} {\bibfnamefont {C.~W.~J.}\ \bibnamefont {Beenakker}},\
  }\href@noop {} {\bibfield  {journal} {\bibinfo  {journal} {Phys. Rev. B}\
  }\textbf {\bibinfo {volume} {83}},\ \bibinfo {pages} {155429} (\bibinfo
  {year} {2011})}\BibitemShut {NoStop}%
\bibitem [{\citenamefont {Akhmerov}\ \emph {et~al.}(2011)\citenamefont
  {Akhmerov}, \citenamefont {Dahlhaus}, \citenamefont {Hassler}, \citenamefont
  {Wimmer},\ and\ \citenamefont {Beenakker}}]{akhmerov2011quantized}%
  \BibitemOpen
  \bibfield  {author} {\bibinfo {author} {\bibfnamefont {A.~R.}\ \bibnamefont
  {Akhmerov}}, \bibinfo {author} {\bibfnamefont {J.~P.}\ \bibnamefont
  {Dahlhaus}}, \bibinfo {author} {\bibfnamefont {F.}~\bibnamefont {Hassler}},
  \bibinfo {author} {\bibfnamefont {M.}~\bibnamefont {Wimmer}}, \ and\ \bibinfo
  {author} {\bibfnamefont {C.~W.~J.}\ \bibnamefont {Beenakker}},\ }\href@noop
  {} {\bibfield  {journal} {\bibinfo  {journal} {Phys. Rev. Lett.}\ }\textbf
  {\bibinfo {volume} {106}},\ \bibinfo {pages} {057001} (\bibinfo {year}
  {2011})}\BibitemShut {NoStop}%
\bibitem [{\citenamefont {Ryu}\ and\ \citenamefont
  {Nomura}(2012)}]{ryu2012disorder}%
  \BibitemOpen
  \bibfield  {author} {\bibinfo {author} {\bibfnamefont {S.}~\bibnamefont
  {Ryu}}\ and\ \bibinfo {author} {\bibfnamefont {K.}~\bibnamefont {Nomura}},\
  }\href@noop {} {\bibfield  {journal} {\bibinfo  {journal} {Phys. Rev. B}\
  }\textbf {\bibinfo {volume} {85}},\ \bibinfo {pages} {155138} (\bibinfo
  {year} {2012})}\BibitemShut {NoStop}%
\bibitem [{\citenamefont {Song}\ \emph {et~al.}(2012)\citenamefont {Song},
  \citenamefont {Liu}, \citenamefont {Jiang}, \citenamefont {Sun},\ and\
  \citenamefont {Xie}}]{song2012dependence}%
  \BibitemOpen
  \bibfield  {author} {\bibinfo {author} {\bibfnamefont {J.}~\bibnamefont
  {Song}}, \bibinfo {author} {\bibfnamefont {H.}~\bibnamefont {Liu}}, \bibinfo
  {author} {\bibfnamefont {H.}~\bibnamefont {Jiang}}, \bibinfo {author}
  {\bibfnamefont {Q.-f.}\ \bibnamefont {Sun}}, \ and\ \bibinfo {author}
  {\bibfnamefont {X.~C.}\ \bibnamefont {Xie}},\ }\href@noop {} {\bibfield
  {journal} {\bibinfo  {journal} {Phys. Rev. B}\ }\textbf {\bibinfo {volume}
  {85}},\ \bibinfo {pages} {195125} (\bibinfo {year} {2012})}\BibitemShut
  {NoStop}%
\bibitem [{\citenamefont {Altland}\ \emph {et~al.}(2014)\citenamefont
  {Altland}, \citenamefont {Bagrets}, \citenamefont {Fritz}, \citenamefont
  {Kamenev},\ and\ \citenamefont {Schmiedt}}]{altland2014quantum}%
  \BibitemOpen
  \bibfield  {author} {\bibinfo {author} {\bibfnamefont {A.}~\bibnamefont
  {Altland}}, \bibinfo {author} {\bibfnamefont {D.}~\bibnamefont {Bagrets}},
  \bibinfo {author} {\bibfnamefont {L.}~\bibnamefont {Fritz}}, \bibinfo
  {author} {\bibfnamefont {A.}~\bibnamefont {Kamenev}}, \ and\ \bibinfo
  {author} {\bibfnamefont {H.}~\bibnamefont {Schmiedt}},\ }\href@noop {}
  {\bibfield  {journal} {\bibinfo  {journal} {Phys. Rev. Lett.}\ }\textbf
  {\bibinfo {volume} {112}},\ \bibinfo {pages} {206602} (\bibinfo {year}
  {2014})}\BibitemShut {NoStop}%
\bibitem [{\citenamefont {Liu}\ \emph {et~al.}(2017)\citenamefont {Liu},
  \citenamefont {Gao}, \citenamefont {Yang},\ and\ \citenamefont
  {Zhang}}]{liu2017disorder}%
  \BibitemOpen
  \bibfield  {author} {\bibinfo {author} {\bibfnamefont {C.}~\bibnamefont
  {Liu}}, \bibinfo {author} {\bibfnamefont {W.}~\bibnamefont {Gao}}, \bibinfo
  {author} {\bibfnamefont {B.}~\bibnamefont {Yang}}, \ and\ \bibinfo {author}
  {\bibfnamefont {S.}~\bibnamefont {Zhang}},\ }\href@noop {} {\bibfield
  {journal} {\bibinfo  {journal} {Phys. Rev. Lett.}\ }\textbf {\bibinfo
  {volume} {119}},\ \bibinfo {pages} {183901} (\bibinfo {year}
  {2017})}\BibitemShut {NoStop}%
\bibitem [{\citenamefont {Meier}\ \emph {et~al.}(2018)\citenamefont {Meier},
  \citenamefont {An}, \citenamefont {Dauphin}, \citenamefont {Maffei},
  \citenamefont {Massignan}, \citenamefont {Hughes},\ and\ \citenamefont
  {Gadway}}]{meier2018observation}%
  \BibitemOpen
  \bibfield  {author} {\bibinfo {author} {\bibfnamefont {E.~J.}\ \bibnamefont
  {Meier}}, \bibinfo {author} {\bibfnamefont {F.~A.}\ \bibnamefont {An}},
  \bibinfo {author} {\bibfnamefont {A.}~\bibnamefont {Dauphin}}, \bibinfo
  {author} {\bibfnamefont {M.}~\bibnamefont {Maffei}}, \bibinfo {author}
  {\bibfnamefont {P.}~\bibnamefont {Massignan}}, \bibinfo {author}
  {\bibfnamefont {T.~L.}\ \bibnamefont {Hughes}}, \ and\ \bibinfo {author}
  {\bibfnamefont {B.}~\bibnamefont {Gadway}},\ }\href@noop {} {\bibfield
  {journal} {\bibinfo  {journal} {Science}\ }\textbf {\bibinfo {volume}
  {362}},\ \bibinfo {pages} {929} (\bibinfo {year} {2018})}\BibitemShut
  {NoStop}%
\bibitem [{\citenamefont {St{\"u}tzer}\ \emph {et~al.}(2018)\citenamefont
  {St{\"u}tzer}, \citenamefont {Plotnik}, \citenamefont {Lumer}, \citenamefont
  {Titum}, \citenamefont {Lindner}, \citenamefont {Segev}, \citenamefont
  {Rechtsman},\ and\ \citenamefont {Szameit}}]{stutzer2018photonic}%
  \BibitemOpen
  \bibfield  {author} {\bibinfo {author} {\bibfnamefont {S.}~\bibnamefont
  {St{\"u}tzer}}, \bibinfo {author} {\bibfnamefont {Y.}~\bibnamefont
  {Plotnik}}, \bibinfo {author} {\bibfnamefont {Y.}~\bibnamefont {Lumer}},
  \bibinfo {author} {\bibfnamefont {P.}~\bibnamefont {Titum}}, \bibinfo
  {author} {\bibfnamefont {N.~H.}\ \bibnamefont {Lindner}}, \bibinfo {author}
  {\bibfnamefont {M.}~\bibnamefont {Segev}}, \bibinfo {author} {\bibfnamefont
  {M.~C.}\ \bibnamefont {Rechtsman}}, \ and\ \bibinfo {author} {\bibfnamefont
  {A.}~\bibnamefont {Szameit}},\ }\href@noop {} {\bibfield  {journal} {\bibinfo
   {journal} {Nature}\ }\textbf {\bibinfo {volume} {560}},\ \bibinfo {pages}
  {461} (\bibinfo {year} {2018})}\BibitemShut {NoStop}%
\bibitem [{\citenamefont {Li}\ \emph {et~al.}(2020)\citenamefont {Li},
  \citenamefont {Fu}, \citenamefont {Hu}, \citenamefont {Li},\ and\
  \citenamefont {Shen}}]{li2020topological}%
  \BibitemOpen
  \bibfield  {author} {\bibinfo {author} {\bibfnamefont {C.-A.}\ \bibnamefont
  {Li}}, \bibinfo {author} {\bibfnamefont {B.}~\bibnamefont {Fu}}, \bibinfo
  {author} {\bibfnamefont {Z.-A.}\ \bibnamefont {Hu}}, \bibinfo {author}
  {\bibfnamefont {J.}~\bibnamefont {Li}}, \ and\ \bibinfo {author}
  {\bibfnamefont {S.-Q.}\ \bibnamefont {Shen}},\ }\href@noop {} {\bibfield
  {journal} {\bibinfo  {journal} {Phys. Rev. Lett.}\ }\textbf {\bibinfo
  {volume} {125}},\ \bibinfo {pages} {166801} (\bibinfo {year}
  {2020})}\BibitemShut {NoStop}%
\bibitem [{\citenamefont {Zhang}\ \emph {et~al.}(2020)\citenamefont {Zhang},
  \citenamefont {Tang}, \citenamefont {Lang}, \citenamefont {Yan},\ and\
  \citenamefont {Zhu}}]{zhang2020non}%
  \BibitemOpen
  \bibfield  {author} {\bibinfo {author} {\bibfnamefont {D.-W.}\ \bibnamefont
  {Zhang}}, \bibinfo {author} {\bibfnamefont {L.-Z.}\ \bibnamefont {Tang}},
  \bibinfo {author} {\bibfnamefont {L.-J.}\ \bibnamefont {Lang}}, \bibinfo
  {author} {\bibfnamefont {H.}~\bibnamefont {Yan}}, \ and\ \bibinfo {author}
  {\bibfnamefont {S.-L.}\ \bibnamefont {Zhu}},\ }\href@noop {} {\bibfield
  {journal} {\bibinfo  {journal} {Sci. China. Phys. Mech.}\ }\textbf {\bibinfo
  {volume} {63}},\ \bibinfo {pages} {267062} (\bibinfo {year}
  {2020})}\BibitemShut {NoStop}%
\bibitem [{\citenamefont {Liu}\ \emph {et~al.}(2020)\citenamefont {Liu},
  \citenamefont {Yang}, \citenamefont {Ren}, \citenamefont {Xue}, \citenamefont
  {Lin}, \citenamefont {Hu}, \citenamefont {Sun}, \citenamefont {Peng},
  \citenamefont {Zhou}, \citenamefont {Chong},\ and\ \citenamefont
  {Zhang}}]{liu2020topological}%
  \BibitemOpen
  \bibfield  {author} {\bibinfo {author} {\bibfnamefont {G.-G.}\ \bibnamefont
  {Liu}}, \bibinfo {author} {\bibfnamefont {Y.}~\bibnamefont {Yang}}, \bibinfo
  {author} {\bibfnamefont {X.}~\bibnamefont {Ren}}, \bibinfo {author}
  {\bibfnamefont {H.}~\bibnamefont {Xue}}, \bibinfo {author} {\bibfnamefont
  {X.}~\bibnamefont {Lin}}, \bibinfo {author} {\bibfnamefont {Y.-H.}\
  \bibnamefont {Hu}}, \bibinfo {author} {\bibfnamefont {H.-x.}\ \bibnamefont
  {Sun}}, \bibinfo {author} {\bibfnamefont {B.}~\bibnamefont {Peng}}, \bibinfo
  {author} {\bibfnamefont {P.}~\bibnamefont {Zhou}}, \bibinfo {author}
  {\bibfnamefont {Y.}~\bibnamefont {Chong}}, \ and\ \bibinfo {author}
  {\bibfnamefont {B.}~\bibnamefont {Zhang}},\ }\href@noop {} {\bibfield
  {journal} {\bibinfo  {journal} {Phys. Rev. Lett.}\ }\textbf {\bibinfo
  {volume} {125}},\ \bibinfo {pages} {133603} (\bibinfo {year}
  {2020})}\BibitemShut {NoStop}%
\bibitem [{\citenamefont {Cui}\ \emph {et~al.}(2022)\citenamefont {Cui},
  \citenamefont {Zhang}, \citenamefont {Zhang},\ and\ \citenamefont
  {Chan}}]{cui2022photonic}%
  \BibitemOpen
  \bibfield  {author} {\bibinfo {author} {\bibfnamefont {X.}~\bibnamefont
  {Cui}}, \bibinfo {author} {\bibfnamefont {R.-Y.}\ \bibnamefont {Zhang}},
  \bibinfo {author} {\bibfnamefont {Z.-Q.}\ \bibnamefont {Zhang}}, \ and\
  \bibinfo {author} {\bibfnamefont {C.~T.}\ \bibnamefont {Chan}},\ }\href@noop
  {} {\bibfield  {journal} {\bibinfo  {journal} {Phys. Rev. Lett.}\ }\textbf
  {\bibinfo {volume} {129}},\ \bibinfo {pages} {043902} (\bibinfo {year}
  {2022})}\BibitemShut {NoStop}%
\bibitem [{\citenamefont {Liu}\ \emph {et~al.}(2021)\citenamefont {Liu},
  \citenamefont {Zhou}, \citenamefont {Wu}, \citenamefont {Zhang},\ and\
  \citenamefont {Jiang}}]{liu2021real}%
  \BibitemOpen
  \bibfield  {author} {\bibinfo {author} {\bibfnamefont {H.}~\bibnamefont
  {Liu}}, \bibinfo {author} {\bibfnamefont {J.-K.}\ \bibnamefont {Zhou}},
  \bibinfo {author} {\bibfnamefont {B.-L.}\ \bibnamefont {Wu}}, \bibinfo
  {author} {\bibfnamefont {Z.-Q.}\ \bibnamefont {Zhang}}, \ and\ \bibinfo
  {author} {\bibfnamefont {H.}~\bibnamefont {Jiang}},\ }\href@noop {}
  {\bibfield  {journal} {\bibinfo  {journal} {Phys. Rev. B}\ }\textbf {\bibinfo
  {volume} {103}},\ \bibinfo {pages} {224203} (\bibinfo {year}
  {2021})}\BibitemShut {NoStop}%
\bibitem [{\citenamefont {Yang}\ \emph
  {et~al.}(2021{\natexlab{a}})\citenamefont {Yang}, \citenamefont {Li},
  \citenamefont {Duan},\ and\ \citenamefont {Xu}}]{yang2021higher}%
  \BibitemOpen
  \bibfield  {author} {\bibinfo {author} {\bibfnamefont {Y.-B.}\ \bibnamefont
  {Yang}}, \bibinfo {author} {\bibfnamefont {K.}~\bibnamefont {Li}}, \bibinfo
  {author} {\bibfnamefont {L.-M.}\ \bibnamefont {Duan}}, \ and\ \bibinfo
  {author} {\bibfnamefont {Y.}~\bibnamefont {Xu}},\ }\href@noop {} {\bibfield
  {journal} {\bibinfo  {journal} {Phys. Rev. B}\ }\textbf {\bibinfo {volume}
  {103}},\ \bibinfo {pages} {085408} (\bibinfo {year}
  {2021}{\natexlab{a}})}\BibitemShut {NoStop}%
\bibitem [{\citenamefont {Shi}\ \emph {et~al.}(2021)\citenamefont {Shi},
  \citenamefont {Kiorpelidis}, \citenamefont {Chaunsali}, \citenamefont
  {Achilleos}, \citenamefont {Theocharis},\ and\ \citenamefont
  {Yang}}]{shi2021disorder}%
  \BibitemOpen
  \bibfield  {author} {\bibinfo {author} {\bibfnamefont {X.}~\bibnamefont
  {Shi}}, \bibinfo {author} {\bibfnamefont {I.}~\bibnamefont {Kiorpelidis}},
  \bibinfo {author} {\bibfnamefont {R.}~\bibnamefont {Chaunsali}}, \bibinfo
  {author} {\bibfnamefont {V.}~\bibnamefont {Achilleos}}, \bibinfo {author}
  {\bibfnamefont {G.}~\bibnamefont {Theocharis}}, \ and\ \bibinfo {author}
  {\bibfnamefont {J.}~\bibnamefont {Yang}},\ }\href@noop {} {\bibfield
  {journal} {\bibinfo  {journal} {Phys. Rev. Research}\ }\textbf {\bibinfo
  {volume} {3}},\ \bibinfo {pages} {033012} (\bibinfo {year}
  {2021})}\BibitemShut {NoStop}%
\bibitem [{\citenamefont {Lin}\ \emph {et~al.}(2022)\citenamefont {Lin},
  \citenamefont {Li}, \citenamefont {Xiao}, \citenamefont {Wang}, \citenamefont
  {Yi},\ and\ \citenamefont {Xue}}]{lin2022observation}%
  \BibitemOpen
  \bibfield  {author} {\bibinfo {author} {\bibfnamefont {Q.}~\bibnamefont
  {Lin}}, \bibinfo {author} {\bibfnamefont {T.}~\bibnamefont {Li}}, \bibinfo
  {author} {\bibfnamefont {L.}~\bibnamefont {Xiao}}, \bibinfo {author}
  {\bibfnamefont {K.}~\bibnamefont {Wang}}, \bibinfo {author} {\bibfnamefont
  {W.}~\bibnamefont {Yi}}, \ and\ \bibinfo {author} {\bibfnamefont
  {P.}~\bibnamefont {Xue}},\ }\href@noop {} {\bibfield  {journal} {\bibinfo
  {journal} {Nat. Commun.}\ }\textbf {\bibinfo {volume} {13}},\ \bibinfo
  {pages} {3229} (\bibinfo {year} {2022})}\BibitemShut {NoStop}%
\bibitem [{\citenamefont {Chen}\ \emph {et~al.}(2019)\citenamefont {Chen},
  \citenamefont {Xu},\ and\ \citenamefont {Zhou}}]{chen2019topological}%
  \BibitemOpen
  \bibfield  {author} {\bibinfo {author} {\bibfnamefont {R.}~\bibnamefont
  {Chen}}, \bibinfo {author} {\bibfnamefont {D.-H.}\ \bibnamefont {Xu}}, \ and\
  \bibinfo {author} {\bibfnamefont {B.}~\bibnamefont {Zhou}},\ }\href@noop {}
  {\bibfield  {journal} {\bibinfo  {journal} {Phys. Rev. B}\ }\textbf {\bibinfo
  {volume} {100}},\ \bibinfo {pages} {115311} (\bibinfo {year}
  {2019})}\BibitemShut {NoStop}%
\bibitem [{\citenamefont {Su}\ \emph {et~al.}(1979)\citenamefont {Su},
  \citenamefont {Schrieffer},\ and\ \citenamefont {Heeger}}]{su1979solitons}%
  \BibitemOpen
  \bibfield  {author} {\bibinfo {author} {\bibfnamefont {W.~P.}\ \bibnamefont
  {Su}}, \bibinfo {author} {\bibfnamefont {J.~R.}\ \bibnamefont {Schrieffer}},
  \ and\ \bibinfo {author} {\bibfnamefont {A.~J.}\ \bibnamefont {Heeger}},\
  }\href@noop {} {\bibfield  {journal} {\bibinfo  {journal} {Phys. Rev. Lett.}\
  }\textbf {\bibinfo {volume} {42}},\ \bibinfo {pages} {1698} (\bibinfo {year}
  {1979})}\BibitemShut {NoStop}%
\bibitem [{\citenamefont {Weaire}\ and\ \citenamefont
  {Thorpe}(1971)}]{weaire1971electronic}%
  \BibitemOpen
  \bibfield  {author} {\bibinfo {author} {\bibfnamefont {D.}~\bibnamefont
  {Weaire}}\ and\ \bibinfo {author} {\bibfnamefont {M.}~\bibnamefont
  {Thorpe}},\ }\href@noop {} {\bibfield  {journal} {\bibinfo  {journal} {Phys.
  Rev. B}\ }\textbf {\bibinfo {volume} {4}},\ \bibinfo {pages} {2508} (\bibinfo
  {year} {1971})}\BibitemShut {NoStop}%
\bibitem [{\citenamefont {Zallen}(1998)}]{zallen2008physics}%
  \BibitemOpen
  \bibfield  {author} {\bibinfo {author} {\bibfnamefont {R.}~\bibnamefont
  {Zallen}},\ }\href@noop {} {\emph {\bibinfo {title} {The physics of amorphous
  solids}}}\ (\bibinfo  {publisher} {John Wiley \& Sons, New York},\ \bibinfo
  {year} {1998})\BibitemShut {NoStop}%
\bibitem [{\citenamefont {Yang}\ \emph
  {et~al.}(2021{\natexlab{b}})\citenamefont {Yang}, \citenamefont {Zhou},
  \citenamefont {Zhu}, \citenamefont {Yuan}, \citenamefont {Chang},
  \citenamefont {Kim}, \citenamefont {Pham}, \citenamefont {Rana},
  \citenamefont {Tian}, \citenamefont {Yao}, \citenamefont {Osher},
  \citenamefont {Schmid}, \citenamefont {Hu}, \citenamefont {Ercius},\ and\
  \citenamefont {Miao}}]{yang2021determining}%
  \BibitemOpen
  \bibfield  {author} {\bibinfo {author} {\bibfnamefont {Y.}~\bibnamefont
  {Yang}}, \bibinfo {author} {\bibfnamefont {J.}~\bibnamefont {Zhou}}, \bibinfo
  {author} {\bibfnamefont {F.}~\bibnamefont {Zhu}}, \bibinfo {author}
  {\bibfnamefont {Y.}~\bibnamefont {Yuan}}, \bibinfo {author} {\bibfnamefont
  {D.~J.}\ \bibnamefont {Chang}}, \bibinfo {author} {\bibfnamefont {D.~S.}\
  \bibnamefont {Kim}}, \bibinfo {author} {\bibfnamefont {M.}~\bibnamefont
  {Pham}}, \bibinfo {author} {\bibfnamefont {A.}~\bibnamefont {Rana}}, \bibinfo
  {author} {\bibfnamefont {X.}~\bibnamefont {Tian}}, \bibinfo {author}
  {\bibfnamefont {Y.}~\bibnamefont {Yao}}, \bibinfo {author} {\bibfnamefont
  {S.~J.}\ \bibnamefont {Osher}}, \bibinfo {author} {\bibfnamefont {A.~K.}\
  \bibnamefont {Schmid}}, \bibinfo {author} {\bibfnamefont {L.}~\bibnamefont
  {Hu}}, \bibinfo {author} {\bibfnamefont {P.}~\bibnamefont {Ercius}}, \ and\
  \bibinfo {author} {\bibfnamefont {J.}~\bibnamefont {Miao}},\ }\href@noop {}
  {\bibfield  {journal} {\bibinfo  {journal} {Nature}\ }\textbf {\bibinfo
  {volume} {592}},\ \bibinfo {pages} {60} (\bibinfo {year}
  {2021}{\natexlab{b}})}\BibitemShut {NoStop}%
\bibitem [{\citenamefont {Focassio}\ \emph
  {et~al.}(2021{\natexlab{a}})\citenamefont {Focassio}, \citenamefont
  {Schleder}, \citenamefont {Costa}, \citenamefont {Fazzio},\ and\
  \citenamefont {Lewenkopf}}]{focassio2021structural}%
  \BibitemOpen
  \bibfield  {author} {\bibinfo {author} {\bibfnamefont {B.}~\bibnamefont
  {Focassio}}, \bibinfo {author} {\bibfnamefont {G.~R.}\ \bibnamefont
  {Schleder}}, \bibinfo {author} {\bibfnamefont {M.}~\bibnamefont {Costa}},
  \bibinfo {author} {\bibfnamefont {A.}~\bibnamefont {Fazzio}}, \ and\ \bibinfo
  {author} {\bibfnamefont {C.}~\bibnamefont {Lewenkopf}},\ }\href@noop {}
  {\bibfield  {journal} {\bibinfo  {journal} {2D Materials}\ }\textbf {\bibinfo
  {volume} {8}},\ \bibinfo {pages} {025032} (\bibinfo {year}
  {2021}{\natexlab{a}})}\BibitemShut {NoStop}%
\bibitem [{\citenamefont {Rechtsman}\ \emph {et~al.}(2011)\citenamefont
  {Rechtsman}, \citenamefont {Szameit}, \citenamefont {Dreisow}, \citenamefont
  {Heinrich}, \citenamefont {Keil}, \citenamefont {Nolte},\ and\ \citenamefont
  {Segev}}]{rechtsman2011amorphous}%
  \BibitemOpen
  \bibfield  {author} {\bibinfo {author} {\bibfnamefont {M.}~\bibnamefont
  {Rechtsman}}, \bibinfo {author} {\bibfnamefont {A.}~\bibnamefont {Szameit}},
  \bibinfo {author} {\bibfnamefont {F.}~\bibnamefont {Dreisow}}, \bibinfo
  {author} {\bibfnamefont {M.}~\bibnamefont {Heinrich}}, \bibinfo {author}
  {\bibfnamefont {R.}~\bibnamefont {Keil}}, \bibinfo {author} {\bibfnamefont
  {S.}~\bibnamefont {Nolte}}, \ and\ \bibinfo {author} {\bibfnamefont
  {M.}~\bibnamefont {Segev}},\ }\href@noop {} {\bibfield  {journal} {\bibinfo
  {journal} {Phys. Rev. Lett.}\ }\textbf {\bibinfo {volume} {106}},\ \bibinfo
  {pages} {193904} (\bibinfo {year} {2011})}\BibitemShut {NoStop}%
\bibitem [{\citenamefont {Mitchell}\ \emph {et~al.}(2018)\citenamefont
  {Mitchell}, \citenamefont {Nash}, \citenamefont {Hexner}, \citenamefont
  {Turner},\ and\ \citenamefont {Irvine}}]{mitchell2018amorphous}%
  \BibitemOpen
  \bibfield  {author} {\bibinfo {author} {\bibfnamefont {N.~P.}\ \bibnamefont
  {Mitchell}}, \bibinfo {author} {\bibfnamefont {L.~M.}\ \bibnamefont {Nash}},
  \bibinfo {author} {\bibfnamefont {D.}~\bibnamefont {Hexner}}, \bibinfo
  {author} {\bibfnamefont {A.~M.}\ \bibnamefont {Turner}}, \ and\ \bibinfo
  {author} {\bibfnamefont {W.~T.}\ \bibnamefont {Irvine}},\ }\href@noop {}
  {\bibfield  {journal} {\bibinfo  {journal} {Nat. Phys.}\ }\textbf {\bibinfo
  {volume} {14}},\ \bibinfo {pages} {380} (\bibinfo {year} {2018})}\BibitemShut
  {NoStop}%
\bibitem [{\citenamefont {P{\"o}yh{\"o}nen}\ \emph {et~al.}(2018)\citenamefont
  {P{\"o}yh{\"o}nen}, \citenamefont {Sahlberg}, \citenamefont {Weststr{\"o}m},\
  and\ \citenamefont {Ojanen}}]{poyhonen2018amorphous}%
  \BibitemOpen
  \bibfield  {author} {\bibinfo {author} {\bibfnamefont {K.}~\bibnamefont
  {P{\"o}yh{\"o}nen}}, \bibinfo {author} {\bibfnamefont {I.}~\bibnamefont
  {Sahlberg}}, \bibinfo {author} {\bibfnamefont {A.}~\bibnamefont
  {Weststr{\"o}m}}, \ and\ \bibinfo {author} {\bibfnamefont {T.}~\bibnamefont
  {Ojanen}},\ }\href@noop {} {\bibfield  {journal} {\bibinfo  {journal} {Nat.
  Commun.}\ }\textbf {\bibinfo {volume} {9}},\ \bibinfo {pages} {2103}
  (\bibinfo {year} {2018})}\BibitemShut {NoStop}%
\bibitem [{\citenamefont {Agarwala}\ and\ \citenamefont
  {Shenoy}(2017)}]{agarwala2019topological}%
  \BibitemOpen
  \bibfield  {author} {\bibinfo {author} {\bibfnamefont {A.}~\bibnamefont
  {Agarwala}}\ and\ \bibinfo {author} {\bibfnamefont {V.~B.}\ \bibnamefont
  {Shenoy}},\ }\href@noop {} {\bibfield  {journal} {\bibinfo  {journal} {Phys.
  Rev. Lett.}\ }\textbf {\bibinfo {volume} {118}},\ \bibinfo {pages} {236402}
  (\bibinfo {year} {2017})}\BibitemShut {NoStop}%
\bibitem [{\citenamefont {Mansha}\ and\ \citenamefont
  {Chong}(2017)}]{mansha2017robust}%
  \BibitemOpen
  \bibfield  {author} {\bibinfo {author} {\bibfnamefont {S.}~\bibnamefont
  {Mansha}}\ and\ \bibinfo {author} {\bibfnamefont {Y.~D.}\ \bibnamefont
  {Chong}},\ }\href@noop {} {\bibfield  {journal} {\bibinfo  {journal} {Phys.
  Rev. B}\ }\textbf {\bibinfo {volume} {96}},\ \bibinfo {pages} {121405(R)}
  (\bibinfo {year} {2017})}\BibitemShut {NoStop}%
\bibitem [{\citenamefont {Costa}\ \emph {et~al.}(2019)\citenamefont {Costa},
  \citenamefont {Schleder}, \citenamefont {Buongiorno~Nardelli}, \citenamefont
  {Lewenkopf},\ and\ \citenamefont {Fazzio}}]{costa2019toward}%
  \BibitemOpen
  \bibfield  {author} {\bibinfo {author} {\bibfnamefont {M.}~\bibnamefont
  {Costa}}, \bibinfo {author} {\bibfnamefont {G.~R.}\ \bibnamefont {Schleder}},
  \bibinfo {author} {\bibfnamefont {M.}~\bibnamefont {Buongiorno~Nardelli}},
  \bibinfo {author} {\bibfnamefont {C.}~\bibnamefont {Lewenkopf}}, \ and\
  \bibinfo {author} {\bibfnamefont {A.}~\bibnamefont {Fazzio}},\ }\href@noop {}
  {\bibfield  {journal} {\bibinfo  {journal} {Nano Lett.}\ }\textbf {\bibinfo
  {volume} {19}},\ \bibinfo {pages} {8941} (\bibinfo {year}
  {2019})}\BibitemShut {NoStop}%
\bibitem [{\citenamefont {Agarwala}\ \emph {et~al.}(2020)\citenamefont
  {Agarwala}, \citenamefont {Juri{\v{c}}i{\'c}},\ and\ \citenamefont
  {Roy}}]{agarwala2020higher}%
  \BibitemOpen
  \bibfield  {author} {\bibinfo {author} {\bibfnamefont {A.}~\bibnamefont
  {Agarwala}}, \bibinfo {author} {\bibfnamefont {V.}~\bibnamefont
  {Juri{\v{c}}i{\'c}}}, \ and\ \bibinfo {author} {\bibfnamefont
  {B.}~\bibnamefont {Roy}},\ }\href@noop {} {\bibfield  {journal} {\bibinfo
  {journal} {Phys. Rev. Research}\ }\textbf {\bibinfo {volume} {2}},\ \bibinfo
  {pages} {012067(R)} (\bibinfo {year} {2020})}\BibitemShut {NoStop}%
\bibitem [{\citenamefont {Spring}\ \emph {et~al.}(2021)\citenamefont {Spring},
  \citenamefont {Akhmerov},\ and\ \citenamefont
  {Varjas}}]{spring2021amorphous}%
  \BibitemOpen
  \bibfield  {author} {\bibinfo {author} {\bibfnamefont {H.}~\bibnamefont
  {Spring}}, \bibinfo {author} {\bibfnamefont {A.~R.}\ \bibnamefont
  {Akhmerov}}, \ and\ \bibinfo {author} {\bibfnamefont {D.}~\bibnamefont
  {Varjas}},\ }\href@noop {} {\bibfield  {journal} {\bibinfo  {journal}
  {SciPost Phys.}\ }\textbf {\bibinfo {volume} {11}},\ \bibinfo {pages} {022}
  (\bibinfo {year} {2021})}\BibitemShut {NoStop}%
\bibitem [{\citenamefont {Focassio}\ \emph
  {et~al.}(2021{\natexlab{b}})\citenamefont {Focassio}, \citenamefont
  {Schleder}, \citenamefont {Crasto~de Lima}, \citenamefont {Lewenkopf},\ and\
  \citenamefont {Fazzio}}]{focassio2021amorphous}%
  \BibitemOpen
  \bibfield  {author} {\bibinfo {author} {\bibfnamefont {B.}~\bibnamefont
  {Focassio}}, \bibinfo {author} {\bibfnamefont {G.~R.}\ \bibnamefont
  {Schleder}}, \bibinfo {author} {\bibfnamefont {F.}~\bibnamefont {Crasto~de
  Lima}}, \bibinfo {author} {\bibfnamefont {C.}~\bibnamefont {Lewenkopf}}, \
  and\ \bibinfo {author} {\bibfnamefont {A.}~\bibnamefont {Fazzio}},\
  }\href@noop {} {\bibfield  {journal} {\bibinfo  {journal} {Phys. Rev. B}\
  }\textbf {\bibinfo {volume} {104}},\ \bibinfo {pages} {214206} (\bibinfo
  {year} {2021}{\natexlab{b}})}\BibitemShut {NoStop}%
\bibitem [{\citenamefont {Yang}\ \emph {et~al.}(2019)\citenamefont {Yang},
  \citenamefont {Qin}, \citenamefont {Deng}, \citenamefont {Duan},\ and\
  \citenamefont {Xu}}]{yang2019topological}%
  \BibitemOpen
  \bibfield  {author} {\bibinfo {author} {\bibfnamefont {Y.-B.}\ \bibnamefont
  {Yang}}, \bibinfo {author} {\bibfnamefont {T.}~\bibnamefont {Qin}}, \bibinfo
  {author} {\bibfnamefont {D.-L.}\ \bibnamefont {Deng}}, \bibinfo {author}
  {\bibfnamefont {L.-M.}\ \bibnamefont {Duan}}, \ and\ \bibinfo {author}
  {\bibfnamefont {Y.}~\bibnamefont {Xu}},\ }\href@noop {} {\bibfield  {journal}
  {\bibinfo  {journal} {Phys. Rev. Lett.}\ }\textbf {\bibinfo {volume} {123}},\
  \bibinfo {pages} {076401} (\bibinfo {year} {2019})}\BibitemShut {NoStop}%
\bibitem [{\citenamefont {Ivaki}\ \emph {et~al.}(2020)\citenamefont {Ivaki},
  \citenamefont {Sahlberg},\ and\ \citenamefont
  {Ojanen}}]{ivaki2020criticality}%
  \BibitemOpen
  \bibfield  {author} {\bibinfo {author} {\bibfnamefont {M.~N.}\ \bibnamefont
  {Ivaki}}, \bibinfo {author} {\bibfnamefont {I.}~\bibnamefont {Sahlberg}}, \
  and\ \bibinfo {author} {\bibfnamefont {T.}~\bibnamefont {Ojanen}},\
  }\href@noop {} {\bibfield  {journal} {\bibinfo  {journal} {Phys. Rev.
  Research}\ }\textbf {\bibinfo {volume} {2}},\ \bibinfo {pages} {043301}
  (\bibinfo {year} {2020})}\BibitemShut {NoStop}%
\bibitem [{\citenamefont {Wang}\ \emph {et~al.}(2022)\citenamefont {Wang},
  \citenamefont {Cheng}, \citenamefont {Liu}, \citenamefont {Liu},\ and\
  \citenamefont {Huang}}]{wang2022structural}%
  \BibitemOpen
  \bibfield  {author} {\bibinfo {author} {\bibfnamefont {C.}~\bibnamefont
  {Wang}}, \bibinfo {author} {\bibfnamefont {T.}~\bibnamefont {Cheng}},
  \bibinfo {author} {\bibfnamefont {Z.}~\bibnamefont {Liu}}, \bibinfo {author}
  {\bibfnamefont {F.}~\bibnamefont {Liu}}, \ and\ \bibinfo {author}
  {\bibfnamefont {H.}~\bibnamefont {Huang}},\ }\href@noop {} {\bibfield
  {journal} {\bibinfo  {journal} {Phys. Rev. Lett.}\ }\textbf {\bibinfo
  {volume} {128}},\ \bibinfo {pages} {056401} (\bibinfo {year}
  {2022})}\BibitemShut {NoStop}%
\bibitem [{\citenamefont {Peng}\ \emph {et~al.}(2022)\citenamefont {Peng},
  \citenamefont {Hua}, \citenamefont {Chen}, \citenamefont {Liu}, \citenamefont
  {Huang},\ and\ \citenamefont {Zhou}}]{peng2022density}%
  \BibitemOpen
  \bibfield  {author} {\bibinfo {author} {\bibfnamefont {T.}~\bibnamefont
  {Peng}}, \bibinfo {author} {\bibfnamefont {C.-B.}\ \bibnamefont {Hua}},
  \bibinfo {author} {\bibfnamefont {R.}~\bibnamefont {Chen}}, \bibinfo {author}
  {\bibfnamefont {Z.-R.}\ \bibnamefont {Liu}}, \bibinfo {author} {\bibfnamefont
  {H.-M.}\ \bibnamefont {Huang}}, \ and\ \bibinfo {author} {\bibfnamefont
  {B.}~\bibnamefont {Zhou}},\ }\href@noop {} {\bibfield  {journal} {\bibinfo
  {journal} {Phys. Rev. B}\ }\textbf {\bibinfo {volume} {106}},\ \bibinfo
  {pages} {125310} (\bibinfo {year} {2022})}\BibitemShut {NoStop}%
\bibitem [{\citenamefont {Wang}\ \emph {et~al.}(2021)\citenamefont {Wang},
  \citenamefont {Yang}, \citenamefont {Dai},\ and\ \citenamefont
  {Xu}}]{wang2021structural}%
  \BibitemOpen
  \bibfield  {author} {\bibinfo {author} {\bibfnamefont {J.-H.}\ \bibnamefont
  {Wang}}, \bibinfo {author} {\bibfnamefont {Y.-B.}\ \bibnamefont {Yang}},
  \bibinfo {author} {\bibfnamefont {N.}~\bibnamefont {Dai}}, \ and\ \bibinfo
  {author} {\bibfnamefont {Y.}~\bibnamefont {Xu}},\ }\href@noop {} {\bibfield
  {journal} {\bibinfo  {journal} {Phys. Rev. Lett.}\ }\textbf {\bibinfo
  {volume} {126}},\ \bibinfo {pages} {206404} (\bibinfo {year}
  {2021})}\BibitemShut {NoStop}%
\bibitem [{\citenamefont {Zhang}\ \emph {et~al.}(2023)\citenamefont {Zhang},
  \citenamefont {Delplace},\ and\ \citenamefont {Fleury}}]{zhang2023anomalous}%
  \BibitemOpen
  \bibfield  {author} {\bibinfo {author} {\bibfnamefont {Z.}~\bibnamefont
  {Zhang}}, \bibinfo {author} {\bibfnamefont {P.}~\bibnamefont {Delplace}}, \
  and\ \bibinfo {author} {\bibfnamefont {R.}~\bibnamefont {Fleury}},\
  }\href@noop {} {\bibfield  {journal} {\bibinfo  {journal} {Sci. Adv.}\
  }\textbf {\bibinfo {volume} {9}},\ \bibinfo {pages} {eadg3186} (\bibinfo
  {year} {2023})}\BibitemShut {NoStop}%
\bibitem [{\citenamefont {Zhou}\ \emph {et~al.}(2020)\citenamefont {Zhou},
  \citenamefont {Liu}, \citenamefont {Ren}, \citenamefont {Yang}, \citenamefont
  {Xue}, \citenamefont {Bi}, \citenamefont {Deng}, \citenamefont {Chong},\ and\
  \citenamefont {Zhang}}]{zhou2020photonic}%
  \BibitemOpen
  \bibfield  {author} {\bibinfo {author} {\bibfnamefont {P.}~\bibnamefont
  {Zhou}}, \bibinfo {author} {\bibfnamefont {G.-G.}\ \bibnamefont {Liu}},
  \bibinfo {author} {\bibfnamefont {X.}~\bibnamefont {Ren}}, \bibinfo {author}
  {\bibfnamefont {Y.}~\bibnamefont {Yang}}, \bibinfo {author} {\bibfnamefont
  {H.}~\bibnamefont {Xue}}, \bibinfo {author} {\bibfnamefont {L.}~\bibnamefont
  {Bi}}, \bibinfo {author} {\bibfnamefont {L.}~\bibnamefont {Deng}}, \bibinfo
  {author} {\bibfnamefont {Y.}~\bibnamefont {Chong}}, \ and\ \bibinfo {author}
  {\bibfnamefont {B.}~\bibnamefont {Zhang}},\ }\href@noop {} {\bibfield
  {journal} {\bibinfo  {journal} {Light Sci. Appl.}\ }\textbf {\bibinfo
  {volume} {9}},\ \bibinfo {pages} {133} (\bibinfo {year} {2020})}\BibitemShut
  {NoStop}%
\bibitem [{\citenamefont {Corbae}\ \emph {et~al.}(2023)\citenamefont {Corbae},
  \citenamefont {Ciocys}, \citenamefont {Varjas}, \citenamefont {Kennedy},
  \citenamefont {Zeltmann}, \citenamefont {Molina-Ruiz}, \citenamefont
  {Griffin}, \citenamefont {Jozwiak}, \citenamefont {Chen}, \citenamefont
  {Wang}, \citenamefont {Minor}, \citenamefont {Scott}, \citenamefont
  {Grushin}, \citenamefont {Lanzara},\ and\ \citenamefont
  {Hellman}}]{corbae2023observation}%
  \BibitemOpen
  \bibfield  {author} {\bibinfo {author} {\bibfnamefont {P.}~\bibnamefont
  {Corbae}}, \bibinfo {author} {\bibfnamefont {S.}~\bibnamefont {Ciocys}},
  \bibinfo {author} {\bibfnamefont {D.}~\bibnamefont {Varjas}}, \bibinfo
  {author} {\bibfnamefont {E.}~\bibnamefont {Kennedy}}, \bibinfo {author}
  {\bibfnamefont {S.}~\bibnamefont {Zeltmann}}, \bibinfo {author}
  {\bibfnamefont {M.}~\bibnamefont {Molina-Ruiz}}, \bibinfo {author}
  {\bibfnamefont {S.~M.}\ \bibnamefont {Griffin}}, \bibinfo {author}
  {\bibfnamefont {C.}~\bibnamefont {Jozwiak}}, \bibinfo {author} {\bibfnamefont
  {Z.}~\bibnamefont {Chen}}, \bibinfo {author} {\bibfnamefont {L.-W.}\
  \bibnamefont {Wang}}, \bibinfo {author} {\bibfnamefont {A.~M.}\ \bibnamefont
  {Minor}}, \bibinfo {author} {\bibfnamefont {M.}~\bibnamefont {Scott}},
  \bibinfo {author} {\bibfnamefont {A.~G.}\ \bibnamefont {Grushin}}, \bibinfo
  {author} {\bibfnamefont {A.}~\bibnamefont {Lanzara}}, \ and\ \bibinfo
  {author} {\bibfnamefont {F.}~\bibnamefont {Hellman}},\ }\href@noop {}
  {\bibfield  {journal} {\bibinfo  {journal} {Nat. Mater.}\ }\textbf {\bibinfo
  {volume} {22}},\ \bibinfo {pages} {200} (\bibinfo {year} {2023})}\BibitemShut
  {NoStop}%
\bibitem [{\citenamefont {Wooten}\ \emph {et~al.}(1985)\citenamefont {Wooten},
  \citenamefont {Winer},\ and\ \citenamefont {Weaire}}]{wooten1985computer}%
  \BibitemOpen
  \bibfield  {author} {\bibinfo {author} {\bibfnamefont {F.}~\bibnamefont
  {Wooten}}, \bibinfo {author} {\bibfnamefont {K.}~\bibnamefont {Winer}}, \
  and\ \bibinfo {author} {\bibfnamefont {D.}~\bibnamefont {Weaire}},\
  }\href@noop {} {\bibfield  {journal} {\bibinfo  {journal} {Phys. Rev. Lett.}\
  }\textbf {\bibinfo {volume} {54}},\ \bibinfo {pages} {1392} (\bibinfo {year}
  {1985})}\BibitemShut {NoStop}%
\bibitem [{\citenamefont {Car}\ and\ \citenamefont
  {Parrinello}(1988)}]{car1988structural}%
  \BibitemOpen
  \bibfield  {author} {\bibinfo {author} {\bibfnamefont {R.}~\bibnamefont
  {Car}}\ and\ \bibinfo {author} {\bibfnamefont {M.}~\bibnamefont
  {Parrinello}},\ }\href@noop {} {\bibfield  {journal} {\bibinfo  {journal}
  {Phys. Rev. Lett.}\ }\textbf {\bibinfo {volume} {60}},\ \bibinfo {pages}
  {204} (\bibinfo {year} {1988})}\BibitemShut {NoStop}%
\bibitem [{\citenamefont {Tu}\ \emph {et~al.}(1998)\citenamefont {Tu},
  \citenamefont {Tersoff}, \citenamefont {Grinstein},\ and\ \citenamefont
  {Vanderbilt}}]{tu1998properties}%
  \BibitemOpen
  \bibfield  {author} {\bibinfo {author} {\bibfnamefont {Y.}~\bibnamefont
  {Tu}}, \bibinfo {author} {\bibfnamefont {J.}~\bibnamefont {Tersoff}},
  \bibinfo {author} {\bibfnamefont {G.}~\bibnamefont {Grinstein}}, \ and\
  \bibinfo {author} {\bibfnamefont {D.}~\bibnamefont {Vanderbilt}},\
  }\href@noop {} {\bibfield  {journal} {\bibinfo  {journal} {Phys. Rev. Lett.}\
  }\textbf {\bibinfo {volume} {81}},\ \bibinfo {pages} {4899} (\bibinfo {year}
  {1998})}\BibitemShut {NoStop}%
\bibitem [{\citenamefont {Hannemann}\ \emph {et~al.}(2004)\citenamefont
  {Hannemann}, \citenamefont {Sch{\"o}n}, \citenamefont {Jansen}, \citenamefont
  {Putz},\ and\ \citenamefont {Lengauer}}]{hannemann2004modeling}%
  \BibitemOpen
  \bibfield  {author} {\bibinfo {author} {\bibfnamefont {A.}~\bibnamefont
  {Hannemann}}, \bibinfo {author} {\bibfnamefont {J.~C.}\ \bibnamefont
  {Sch{\"o}n}}, \bibinfo {author} {\bibfnamefont {M.}~\bibnamefont {Jansen}},
  \bibinfo {author} {\bibfnamefont {H.}~\bibnamefont {Putz}}, \ and\ \bibinfo
  {author} {\bibfnamefont {T.}~\bibnamefont {Lengauer}},\ }\href@noop {}
  {\bibfield  {journal} {\bibinfo  {journal} {Phys. Rev. B}\ }\textbf {\bibinfo
  {volume} {70}},\ \bibinfo {pages} {144201} (\bibinfo {year}
  {2004})}\BibitemShut {NoStop}%
\bibitem [{\citenamefont {Drabold}(2009)}]{drabold2009topics}%
  \BibitemOpen
  \bibfield  {author} {\bibinfo {author} {\bibfnamefont {D.}~\bibnamefont
  {Drabold}},\ }\href@noop {} {\bibfield  {journal} {\bibinfo  {journal} {Eur.
  Phys. J. B}\ }\textbf {\bibinfo {volume} {68}},\ \bibinfo {pages} {1}
  (\bibinfo {year} {2009})}\BibitemShut {NoStop}%
\bibitem [{\citenamefont {Youn}\ \emph {et~al.}(2014)\citenamefont {Youn},
  \citenamefont {Kang},\ and\ \citenamefont {Han}}]{youn2014efficient}%
  \BibitemOpen
  \bibfield  {author} {\bibinfo {author} {\bibfnamefont {Y.}~\bibnamefont
  {Youn}}, \bibinfo {author} {\bibfnamefont {Y.}~\bibnamefont {Kang}}, \ and\
  \bibinfo {author} {\bibfnamefont {S.}~\bibnamefont {Han}},\ }\href@noop {}
  {\bibfield  {journal} {\bibinfo  {journal} {Comput. Mater. Sci.}\ }\textbf
  {\bibinfo {volume} {95}},\ \bibinfo {pages} {256} (\bibinfo {year}
  {2014})}\BibitemShut {NoStop}%
\bibitem [{\citenamefont {Duwez}\ \emph {et~al.}(1960)\citenamefont {Duwez},
  \citenamefont {Willens},\ and\ \citenamefont
  {Klement~Jr}}]{duwez1960continuous}%
  \BibitemOpen
  \bibfield  {author} {\bibinfo {author} {\bibfnamefont {P.}~\bibnamefont
  {Duwez}}, \bibinfo {author} {\bibfnamefont {R.}~\bibnamefont {Willens}}, \
  and\ \bibinfo {author} {\bibfnamefont {W.}~\bibnamefont {Klement~Jr}},\
  }\href@noop {} {\bibfield  {journal} {\bibinfo  {journal} {J. Appl. Phys.}\
  }\textbf {\bibinfo {volume} {31}},\ \bibinfo {pages} {1136} (\bibinfo {year}
  {1960})}\BibitemShut {NoStop}%
\bibitem [{\citenamefont {Frechette}(1958)}]{koski1961non}%
  \BibitemOpen
  \bibfield  {author} {\bibinfo {author} {\bibfnamefont {V.}~\bibnamefont
  {Frechette}},\ }\href@noop {} {\emph {\bibinfo {title} {Non-Crystalline
  Solids}}}\ (\bibinfo  {publisher} {Wiley, New York},\ \bibinfo {year}
  {1958})\BibitemShut {NoStop}%
\bibitem [{\citenamefont {Wang}\ \emph {et~al.}(2016)\citenamefont {Wang},
  \citenamefont {Jin},\ and\ \citenamefont {Liu}}]{wang2016quantum}%
  \BibitemOpen
  \bibfield  {author} {\bibinfo {author} {\bibfnamefont {Z.}~\bibnamefont
  {Wang}}, \bibinfo {author} {\bibfnamefont {K.-H.}\ \bibnamefont {Jin}}, \
  and\ \bibinfo {author} {\bibfnamefont {F.}~\bibnamefont {Liu}},\ }\href@noop
  {} {\bibfield  {journal} {\bibinfo  {journal} {Nat. Commun.}\ }\textbf
  {\bibinfo {volume} {7}},\ \bibinfo {pages} {12746} (\bibinfo {year}
  {2016})}\BibitemShut {NoStop}%
\bibitem [{\citenamefont {Huang}\ and\ \citenamefont
  {Liu}(2018{\natexlab{a}})}]{huang2018quantum}%
  \BibitemOpen
  \bibfield  {author} {\bibinfo {author} {\bibfnamefont {H.}~\bibnamefont
  {Huang}}\ and\ \bibinfo {author} {\bibfnamefont {F.}~\bibnamefont {Liu}},\
  }\href@noop {} {\bibfield  {journal} {\bibinfo  {journal} {Phys. Rev. Lett.}\
  }\textbf {\bibinfo {volume} {121}},\ \bibinfo {pages} {126401} (\bibinfo
  {year} {2018}{\natexlab{a}})}\BibitemShut {NoStop}%
\bibitem [{\citenamefont {Slater}\ and\ \citenamefont
  {Koster}(1954)}]{slater1954simplified}%
  \BibitemOpen
  \bibfield  {author} {\bibinfo {author} {\bibfnamefont {J.~C.}\ \bibnamefont
  {Slater}}\ and\ \bibinfo {author} {\bibfnamefont {G.~F.}\ \bibnamefont
  {Koster}},\ }\href@noop {} {\bibfield  {journal} {\bibinfo  {journal} {Phys.
  Rev.}\ }\textbf {\bibinfo {volume} {94}},\ \bibinfo {pages} {1498} (\bibinfo
  {year} {1954})}\BibitemShut {NoStop}%
\bibitem [{\citenamefont {Harrison}(1989)}]{harrison2012electronic}%
  \BibitemOpen
  \bibfield  {author} {\bibinfo {author} {\bibfnamefont {W.~A.}\ \bibnamefont
  {Harrison}},\ }\href@noop {} {\emph {\bibinfo {title} {Electronic structure
  and the properties of solids: the physics of the chemical bond}}}\ (\bibinfo
  {publisher} {Dover Publications},\ \bibinfo {year} {1989})\BibitemShut
  {NoStop}%
\bibitem [{\citenamefont {Huang}\ and\ \citenamefont
  {Liu}(2018{\natexlab{b}})}]{huang2018theory}%
  \BibitemOpen
  \bibfield  {author} {\bibinfo {author} {\bibfnamefont {H.}~\bibnamefont
  {Huang}}\ and\ \bibinfo {author} {\bibfnamefont {F.}~\bibnamefont {Liu}},\
  }\href@noop {} {\bibfield  {journal} {\bibinfo  {journal} {Phys. Rev. B}\
  }\textbf {\bibinfo {volume} {98}},\ \bibinfo {pages} {125130} (\bibinfo
  {year} {2018}{\natexlab{b}})}\BibitemShut {NoStop}%
\bibitem [{sup()}]{suppsee}%
  \BibitemOpen
  \href@noop {} {}\bibinfo {howpublished} {See Supplemental Material at
  {http://link.aps.org/supplemental/XXX} for more details about the theoretical
  analysis and numerical results, which includes Refs.
  \cite{koski1961non,huang2018quantum,huang2018theory,datta1997electronic,xing2011topological,zhang2014universal,khomyakov2005conductance,wang2009relation,zhang2012first,groth2009theory,wang2016quantum,wang2022structural,cheng2021antichiral,toniolo2018time,chen2006universal,sheng2006introduction}}\BibitemShut
  {NoStop}%
\bibitem [{\citenamefont {Datta}(1997)}]{datta1997electronic}%
  \BibitemOpen
  \bibfield  {author} {\bibinfo {author} {\bibfnamefont {S.}~\bibnamefont
  {Datta}},\ }\href@noop {} {\emph {\bibinfo {title} {Electronic transport in
  mesoscopic systems}}}\ (\bibinfo  {publisher} {Cambridge university},\
  \bibinfo {year} {1997})\BibitemShut {NoStop}%
\bibitem [{\citenamefont {Khomyakov}\ \emph {et~al.}(2005)\citenamefont
  {Khomyakov}, \citenamefont {Brocks}, \citenamefont {Karpan}, \citenamefont
  {Zwierzycki},\ and\ \citenamefont {Kelly}}]{khomyakov2005conductance}%
  \BibitemOpen
  \bibfield  {author} {\bibinfo {author} {\bibfnamefont {P.~A.}\ \bibnamefont
  {Khomyakov}}, \bibinfo {author} {\bibfnamefont {G.}~\bibnamefont {Brocks}},
  \bibinfo {author} {\bibfnamefont {V.}~\bibnamefont {Karpan}}, \bibinfo
  {author} {\bibfnamefont {M.}~\bibnamefont {Zwierzycki}}, \ and\ \bibinfo
  {author} {\bibfnamefont {P.~J.}\ \bibnamefont {Kelly}},\ }\href@noop {}
  {\bibfield  {journal} {\bibinfo  {journal} {Phys. Rev. B}\ }\textbf {\bibinfo
  {volume} {72}},\ \bibinfo {pages} {035450} (\bibinfo {year}
  {2005})}\BibitemShut {NoStop}%
\bibitem [{\citenamefont {Wang}\ and\ \citenamefont
  {Guo}(2009)}]{wang2009relation}%
  \BibitemOpen
  \bibfield  {author} {\bibinfo {author} {\bibfnamefont {J.}~\bibnamefont
  {Wang}}\ and\ \bibinfo {author} {\bibfnamefont {H.}~\bibnamefont {Guo}},\
  }\href@noop {} {\bibfield  {journal} {\bibinfo  {journal} {Phys. Rev. B}\
  }\textbf {\bibinfo {volume} {79}},\ \bibinfo {pages} {045119} (\bibinfo
  {year} {2009})}\BibitemShut {NoStop}%
\bibitem [{\citenamefont {Zhang}\ \emph {et~al.}(2012)\citenamefont {Zhang},
  \citenamefont {Xing},\ and\ \citenamefont {Wang}}]{zhang2012first}%
  \BibitemOpen
  \bibfield  {author} {\bibinfo {author} {\bibfnamefont {L.}~\bibnamefont
  {Zhang}}, \bibinfo {author} {\bibfnamefont {Y.}~\bibnamefont {Xing}}, \ and\
  \bibinfo {author} {\bibfnamefont {J.}~\bibnamefont {Wang}},\ }\href@noop {}
  {\bibfield  {journal} {\bibinfo  {journal} {Phys. Rev. B}\ }\textbf {\bibinfo
  {volume} {86}},\ \bibinfo {pages} {155438} (\bibinfo {year}
  {2012})}\BibitemShut {NoStop}%
\bibitem [{\citenamefont {Zhang}\ \emph {et~al.}(2014)\citenamefont {Zhang},
  \citenamefont {Zhuang}, \citenamefont {Xing}, \citenamefont {Li},
  \citenamefont {Wang},\ and\ \citenamefont {Guo}}]{zhang2014universal}%
  \BibitemOpen
  \bibfield  {author} {\bibinfo {author} {\bibfnamefont {L.}~\bibnamefont
  {Zhang}}, \bibinfo {author} {\bibfnamefont {J.}~\bibnamefont {Zhuang}},
  \bibinfo {author} {\bibfnamefont {Y.}~\bibnamefont {Xing}}, \bibinfo {author}
  {\bibfnamefont {J.}~\bibnamefont {Li}}, \bibinfo {author} {\bibfnamefont
  {J.}~\bibnamefont {Wang}}, \ and\ \bibinfo {author} {\bibfnamefont
  {H.}~\bibnamefont {Guo}},\ }\href@noop {} {\bibfield  {journal} {\bibinfo
  {journal} {Phys. Rev. B}\ }\textbf {\bibinfo {volume} {89}},\ \bibinfo
  {pages} {245107} (\bibinfo {year} {2014})}\BibitemShut {NoStop}%
\bibitem [{\citenamefont {Cheng}\ \emph {et~al.}(2021)\citenamefont {Cheng},
  \citenamefont {Chen}, \citenamefont {Zhang}, \citenamefont {Xiao},\ and\
  \citenamefont {Jia}}]{cheng2021antichiral}%
  \BibitemOpen
  \bibfield  {author} {\bibinfo {author} {\bibfnamefont {X.}~\bibnamefont
  {Cheng}}, \bibinfo {author} {\bibfnamefont {J.}~\bibnamefont {Chen}},
  \bibinfo {author} {\bibfnamefont {L.}~\bibnamefont {Zhang}}, \bibinfo
  {author} {\bibfnamefont {L.}~\bibnamefont {Xiao}}, \ and\ \bibinfo {author}
  {\bibfnamefont {S.}~\bibnamefont {Jia}},\ }\href@noop {} {\bibfield
  {journal} {\bibinfo  {journal} {Phys. Rev. B}\ }\textbf {\bibinfo {volume}
  {104}},\ \bibinfo {pages} {L081401} (\bibinfo {year} {2021})}\BibitemShut
  {NoStop}%
\bibitem [{\citenamefont {Toniolo}(2018)}]{toniolo2018time}%
  \BibitemOpen
  \bibfield  {author} {\bibinfo {author} {\bibfnamefont {D.}~\bibnamefont
  {Toniolo}},\ }\href@noop {} {\bibfield  {journal} {\bibinfo  {journal} {Phys.
  Rev. B}\ }\textbf {\bibinfo {volume} {98}},\ \bibinfo {pages} {235425}
  (\bibinfo {year} {2018})}\BibitemShut {NoStop}%
\bibitem [{\citenamefont {Chen}\ \emph {et~al.}(2006)\citenamefont {Chen},
  \citenamefont {L{\'o}pez}, \citenamefont {Havlin},\ and\ \citenamefont
  {Stanley}}]{chen2006universal}%
  \BibitemOpen
  \bibfield  {author} {\bibinfo {author} {\bibfnamefont {Y.}~\bibnamefont
  {Chen}}, \bibinfo {author} {\bibfnamefont {E.}~\bibnamefont {L{\'o}pez}},
  \bibinfo {author} {\bibfnamefont {S.}~\bibnamefont {Havlin}}, \ and\ \bibinfo
  {author} {\bibfnamefont {H.~E.}\ \bibnamefont {Stanley}},\ }\href@noop {}
  {\bibfield  {journal} {\bibinfo  {journal} {Phys. Rev. Lett.}\ }\textbf
  {\bibinfo {volume} {96}},\ \bibinfo {pages} {068702} (\bibinfo {year}
  {2006})}\BibitemShut {NoStop}%
\bibitem [{\citenamefont {Sheng}(2006)}]{sheng2006introduction}%
  \BibitemOpen
  \bibfield  {author} {\bibinfo {author} {\bibfnamefont {P.}~\bibnamefont
  {Sheng}},\ }\href@noop {} {\emph {\bibinfo {title} {Introduction to wave
  scattering, localization and mesoscopic phenomena}}}\ (\bibinfo  {publisher}
  {Springer, Berlin},\ \bibinfo {year} {2006})\BibitemShut {NoStop}%
\end{thebibliography}%

\end{document}